\documentclass[twocolumn,appendixfloats]{aastex6}
\slugcomment{{\sc Accepted to AJ:} 5/3/2016}
\usepackage{graphicx}
\usepackage{natbib}
\usepackage{latexsym}
\usepackage{amssymb}
\usepackage{longtable}
\usepackage{amsmath}
\usepackage{url}
\def\msun{\ifmmode {\rm\,M_\odot}\else ${\rm\,M_\odot}$\fi}
\def\Msun{\ifmmode {\rm\,\it{M_\odot}}\else ${\rm\,M_\odot}$\fi}
\def\lsun{\ifmmode {\rm\,L_\odot}\else ${\rm\,L_\odot}$\fi}
\def\Lsun{\ifmmode {\rm\,\it{L_\odot}}\else ${\rm\,L_\odot}$\fi}
\def\rsun{\ifmmode {\rm\,R_\odot}\else ${\rm\,R_\odot}$\fi}
\def\Rsun{\ifmmode {\rm\,\it{R_\odot}}\else ${\rm\,R_\odot}$\fi}
\def\Tsun{\ifmmode {\rm\,T_\odot}\else ${\rm\,T_\odot}$\fi}
\def\arcsec{\ifmmode {^{\prime\prime}}\else $^{\prime\prime}$\fi}
\def\asec{\ifmmode {^{\prime\prime}}\else $^{\prime\prime}$\fi}
\def\arcmin{\ifmmode {^{\prime}}\else $^{\prime}$\fi}
\def\amin{\ifmmode {^{\prime}}\else $^{\prime}$\fi}
\def\simlt{\mathrel{\spose{\lower 3pt\hbox{$\mathchar"218$}}
     \raise 2.0pt\hbox{$\mathchar"13C$}}}
\def\simgt{\mathrel{\spose{\lower 3pt\hbox{$\mathchar"218$}}
\     \raise 2.0pt\hbox{$\mathchar"13E$}}}

\def\wca{W$_{CaH}$}
\def\wha{W$_{H\alpha}$}

\begin{document}

\title{Variation in the pre-transit Balmer line signal around the hot Jupiter HD 189733b}

\author{P. Wilson Cauley and Seth Redfield}
\email{pcauley@wesleyan.edu}
\affil{Wesleyan University\\
Astronomy Department, Van Vleck Observatory, 96 Foss Hill Dr., Middletown, CT 06459}

\author{Adam G. Jensen}
\affil{University of Nebraska-Kearney\\
Department of Physics \& Physical Science, 24011 11th Avenue, Kearney, NE 68849}

\author{Travis Barman}
\affil{University of Arizona\\
Department of Planetary Sciences and Lunar and Planetary Laboratory, 1629 E University Blvd., Tucson, AZ 85721}

%\author{Michael Endl}
%\affil{The University of Texas at Austin\\
%Department of Astronomy and McDonald Observatory, 2515 Speedway, C1400, Austin, TX 78712}

\begin{abstract} 

As followup to our recent detection of a pre-transit
signal around HD 189733 b, we obtained full pre-transit phase coverage 
of a single planetary transit. The pre-transit
signal is again detected in the Balmer lines but with variable strength and timing, suggesting that the bow shock
geometry reported in our previous work does not describe the signal from the latest transit. We also demonstrate
the use of the \ion{Ca}{2} H and K residual core flux as a proxy for the stellar activity level throughout the transit.
A moderate trend is found between the pre-transit absorption signal in the 2013 data and the \ion{Ca}{2} H flux. 
This suggests that some of the 2013 pre-transit hydrogen absorption can be attributed to varying stellar activity levels.
A very weak correlation is found between the \ion{Ca}{2} H core flux and the Balmer
line absorption in the 2015 transit, hinting at a smaller contribution from stellar activity compared to the
2013 transit. We simulate how varying stellar
activity levels can produce changes in the Balmer line transmission spectra. These simulations
show that the strength of the 2013 and 2015 pre-transit signals can be reproduced by stellar variability.
If the pre-transit signature is attributed to circumplanetary material, its evolution in time can be described
by accretion clumps spiraling towards the star, although this interpretation has serious limitations. 
Further high-cadence monitoring at H$\alpha$ is necessary to distinguish between true absorption by transiting 
material and short-term variations in the stellar activity level.

\end{abstract}

\keywords{}

\section{INTRODUCTION}
\label{sec:intro}

Hot exoplanets, i.e., exoplanets with orbital periods of approximately a few days, are unique
laboratories for the study of star-planet interactions (hereafter SPIs). The large amounts of
stellar UV and EUV radiation that these planets receive heats the planetary atmosphere and can
result in significant photoevaporative mass loss across a broad range of planetary masses
\citep[e.g.][]{vidal,murray09,desetangs,owen12,lopez13,ehren15}. If the mass loss is significant and
the planet overflows its Roche lobe, the outflowing planetary material can accrete onto the star
\citep{lai,matsakos,pillitteri15}.  The short orbital distances of these planets can also result in
large tidal interactions, generating enhanced activity levels on the star that are directly related
to the orbital phase of the planet \citep{cuntz,shkolnik08,pillitteri15}. Direct magnetic
interactions via reconnection events or magnetic torques are also possible and can affect the
orbital and spin evolution of the planet and star \citep{lanza10,cohen10,strugarek}. Understanding
these interactions is important in order to move towards a full characterization of exoplanets and a
complete knowledge of their evolution under extreme conditions.  

Two particularly interesting classes of SPIs are the direct interaction of escaping planetary
material (i.e., the planetary wind) and the stellar wind and the interaction of the planet's
magnetosphere with the stellar wind. Both scenarios are capable of producing bow shocks at the
interaction region which results in a significant density enhancement of the shocked material
\citep{vidotto10,bisikalo13,matsakos}. For parameters relevant to most hot planet systems, the
interaction region in both cases occurs ahead of the planet in its orbit. Observationally this can
produce a pre-transit signature (for transiting or very nearly transiting planets) if the material
ahead of the planet is sufficiently opaque to produce measurable absorption
\citep{vidotto10,vidotto11,llama11}. In the case of a bow shock mediated by the planet's
magnetosphere, it is possible to estimate the strength of the field given some assumptions about the
stellar wind \citep{vidotto10,llama13,cauley15}. While estimates of this sort are rough, this method
of measuring exoplanet magnetic fields may be worth pursuing considering the challenges of directly
detecting emission from electrons in the planetary magnetosphere and the current lack of such
confirmed detections \citep[e.g.,][]{murphy15,vidotto15}.  

A third scenario capable of producing pre-transit absorption is the loss of planetary material that
then accretes onto the central star. Accretion streams or time-variable blobs are predicted from
some simulations of hot planet mass loss \citep{cohen11,matsakos} and have been suggested as the cause of
observed UV flares for HD 189733 b \citep{pillitteri15}. Accretion streams or blobs can be viewed in
transmission if the optical depth is sufficiently high in the transition of interest. 

Besides material transiting ahead of the planet, changes in the stellar activity level can be also
produce observable changes in the pre-transit transmission spectrum: if spectra from a period of low
activity are compared to spectra from a period of higher activity, the relative difference can
produce features similar to what is seen in tranmission spectrum absorption lines. HD 189733 is an
active early K-star \citep{boisse09} and, especially for the Balmer line analysis presented here,
contributions from the stellar chromosphere may be non-negligible. These contributions need to be
taken into account when interpreting transmission spectra of lines produced in stellar active
regions.

Evidence for pre-transit material has now been observed in a handful of hot planet systems, namely
WASP-12 b \citep{fossati}, HD 189733 b \citep{benjaffel,bourrier13,cauley15}, and GJ 436 b
\citep{ehren15} \citep[see Section 1 of][for a brief overview]{cauley15}. Most recently,
\citet{ehren15} reported a large pre-transit absorption signature, as well as enhanced in-transit
absorption, in Lyman-$\alpha$ around the hot Neptune GJ 436 b. They model the early ingress as an
extended cloud of hydrogen which has escaped from the planet. The escaping material is subject
to radiation pressure from the star, which in the case of GJ 436 is too weak to overcome the
stellar gravity and instead acts as a radiative braking agent on the gas \citep{bourrier15a}. 
We note that \citep{ehren15} do not include the interaction of the
planetary wind with the stellar wind. In \defcitealias{cauley15}{Paper I}\citet[][hereafter Paper
I]{cauley15} we describe a pre-transit absorption measurement in the Balmer lines for the hot
Jupiter HD 189733 b \citep{bouchy}. We showed that the absorption strength and time series evolution
of the absorption is consistent with the geometry of a thin bow shock at $\sim$13 $R_p$ ahead of the
planet. There are difficulties, however, in maintaining a population of hot neutral hydrogen at
large distances from the planet (Huang \& Christie, private communication). Furthermore, it is
unlikely that the densities produced by a compressed stellar wind at typical hot Jupiter orbital
distances are high enough in most atomic species to generate the necessary opacity, although
escaping planetary material trapped in the magnetosphere may be sufficient \citep{turner16}.

Regardless of their precise interpretation, it appears that pre-transit absorption signals should be
fairly common for hot planets. However, detections of these phenomena seem to be limited to specific
atomic transitions; they have not yet been detected using near-UV broadband methods
\citep[e.g.,][]{turner13,bento14,zellem15} nor, to the best of our knowledge, optical photometry.
High-resolution spectroscopic observations are expensive and are limited to the brightest transiting
targets. As a result, the detection and characterization of these signals, including their
variability, will require a significant observational investment.

In this paper we present followup observations to the 2013 transit of HD 189733 b presented in
\citetalias{cauley15}.  Our observations of the 2013 transit missed a large portion of the
pre-transit phase due to an observing strategy that was not designed to monitor pre-transit signals.
The 2015 transit observations provide complete coverage of the $\sim$4 hours of visible pre-transit
phase. The instrumental setup and experimental design are almost identical to the 2013 transit,
facilitating a direct comparison between the two measurement sets. In \autoref{sec:observations} we
briefly describe the transit observations and data reduction. \autoref{sec:transspec} details the
construction of the Balmer line, \ion{Na}{1} D lines, and \ion{Mg}{1} 5184 \AA\ transmission spectra
and the line absorption time series for each line. We present an analysis of the \ion{Ca}{2} H and K
lines, which serve as a proxy of the stellar activity level, in \autoref{sec:calcium2}.  The
residual core flux analysis in \autoref{sec:calcium2} differs from the Mt. Wilson $S_{HK}$ analysis
presented in \citetalias{cauley15} and provides a more precise estimate of the relative stellar
activity level. Contrary to the $S_{HK}$ index analysis presented in \citetalias{cauley15}, the core
flux analysis shows that there is some relationship between the stellar activity level and the
measured absorption.  We present possible model scenarios to describe the new pre-transit absorption
in \autoref{sec:model}.  We will present a detailed analysis of the in-transit absorption in a
forthcoming paper.  In \autoref{sec:summary} we give a general summary of the results and comment on
future work that is required to clarify the nature of the pre-transit signature.

\section{Observations and data reduction}
\label{sec:observations}

A single transit epoch of HD 189733 b was obtained on the night of August, 4 2015 using HiRES on
Keck I \citep{vogt}. The instrument setup and experimental design were identical to the 2013 transit
observations described in \citetalias{cauley15}. We briefly review them here. The B2 decker was used
and the resolving power of the observations is $R$$\sim$68,000 at H$\alpha$. Exposure times were 5
minutes for all observations resulting in an average signal-to-noise of 450 for the extracted
spectra at H$\alpha$, 200 at H$\beta$, and 150 at H$\gamma$.  The data were reduced using the HiRES
Redux package written by Jason X. Prochaska\footnote{http://www.ucolick.org/$\sim$xavier/HIRedux/}.
All standard reduction steps were taken and the barycentric velocity of the observatory is removed,
along with the radial velocity of the HD 189733 system which we take to be $-2.24$ km s$^{-1}$
\citep{digloria15}, leaving the spectra in the rest frame of the star. Wavelength solutions are
performed on Th-Ar exposures taken at the beginning and end of the night.  Telluric absorption is
removed from the H$\alpha$ and \ion{Na}{1} spectra using the telluric fitting program Molecfit
\citep{kausch}.  A telluric standard was observed at the beginning of the night. The telluric model
is first applied to the standard and then scaled and shifted to fit each individual observation
using a least-squares minimization routine.

\section{Transmission spectra}
\label{sec:transspec}

\begin{figure}[htbp]
   \centering
   \includegraphics[scale=.5,clip,trim=50mm 25mm 50mm 40mm,angle=0]{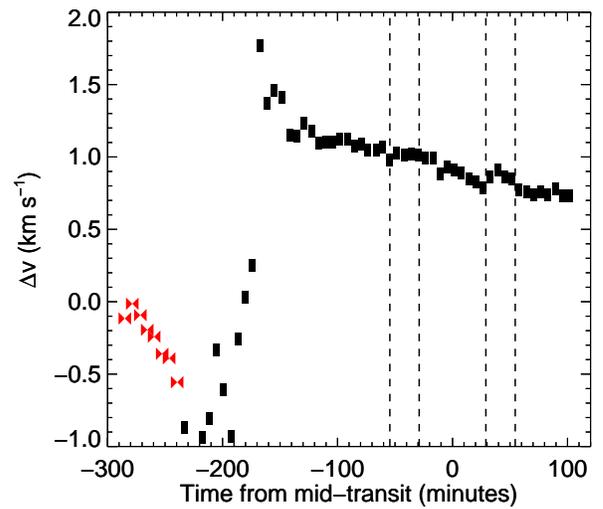} 
   \figcaption{Velocity corrections applied to the H$\alpha$ order, relative to the first spectrum (not shown) for each observation.
   The comparison spectrum corrections are shown with red bow ties. 
   The abrupt changes at $-200$ and $-160$ minutes are due to small guiding errors that resulted in the telescope
   being re-centered on the target.\label{fig:halphavcorrs}}
\end{figure}

The transmission spectrum is defined here as 

\begin{equation}\label{eq:strans}
S_T=\frac{F_{i}}{F_{out}}-1
\end{equation}

\noindent where $F_i$ is a single observation and $F_{out}$ is the master comparison spectrum. The
master comparison spectrum is a weighted average of a set of spectra that are, ideally, free of any
absorption contamination, i.e., they represent the pure stellar spectrum. In order for each $F_i$ to
be compared to $F_{out}$, the spectra need to be normalized and aligned in wavelength space. The
required constant wavelength shifts from one observation to the next, which are calculated by
cross-correlating a large number of metal lines in the order of interest, are of the order 0.01-0.02
\AA\ or 0.5-1.0 km s$^{-1}$. The corrections applied to the H$\alpha$ order, relative to the first
observed spectrum, are shown in \autoref{fig:halphavcorrs}. Higher order functions, e.g., linear or
spline fits, are not necessary to align the spectra. After the spectra are aligned and divided by
$F_{out}$ they are renormalized to remove any residual slope in the continuum.

We note that HiRES is a very stable spectrograph and does not
produce large instrumental effects on the observed spectrum. This is important for having confidence
in the transmission spectrum signal as being due to real changes in the star-planet system and not
from instrument and telescope systematics. Furthermore, any systematic shifting of the spectrum on
the detector or changes in resolution would uniformly affect all stellar lines in a single order. This is
never observed: all of the transmission line profiles reported here occur only in the cores of the stellar lines of
interest and not in the plethora of other stellar lines present in every order. Thus we are confident all of the
reported transmission signals are intrinsic to the star-planet system. 

In order to create the transmission spectrum, the comparison spectra used to generate the master
comparison spectrum must be selected. Since all of the
absorption is measured relative to the master comparison spectrum, choosing different spectra
will increase or decrease the measured absorption values. Thus the true zero-point is difficult to
determine for the small window of time that a single night of observing provides. We have chosen to use
the eight spectra from $t-t_{mid}=-278$ minutes to $t-t_{mid}=-233$ as the comparison spectra for the analysis
presented below in the rest of \autoref{sec:transspec}. These spectra are chosen since they are the earliest spectra
obtained during the night and are farthest from the planet transit in time. If pre-transit absorption is due to
extended structures around the planet, the farther the observations occur from the planet transit the less likely
those observations are to contain a signal from optically thick circumplanetary material. Of course, this does not
guarantee that the spectra are uncontaminated by such material. Our choice of comparison spectra and the
consequences for interpreting the absorption measurements will be discussed in \autoref{sec:compspec}.

\begin{figure*}[htbp]
   \centering
   \includegraphics[scale=.68,clip,trim=5mm 25mm 5mm 30mm,angle=0]{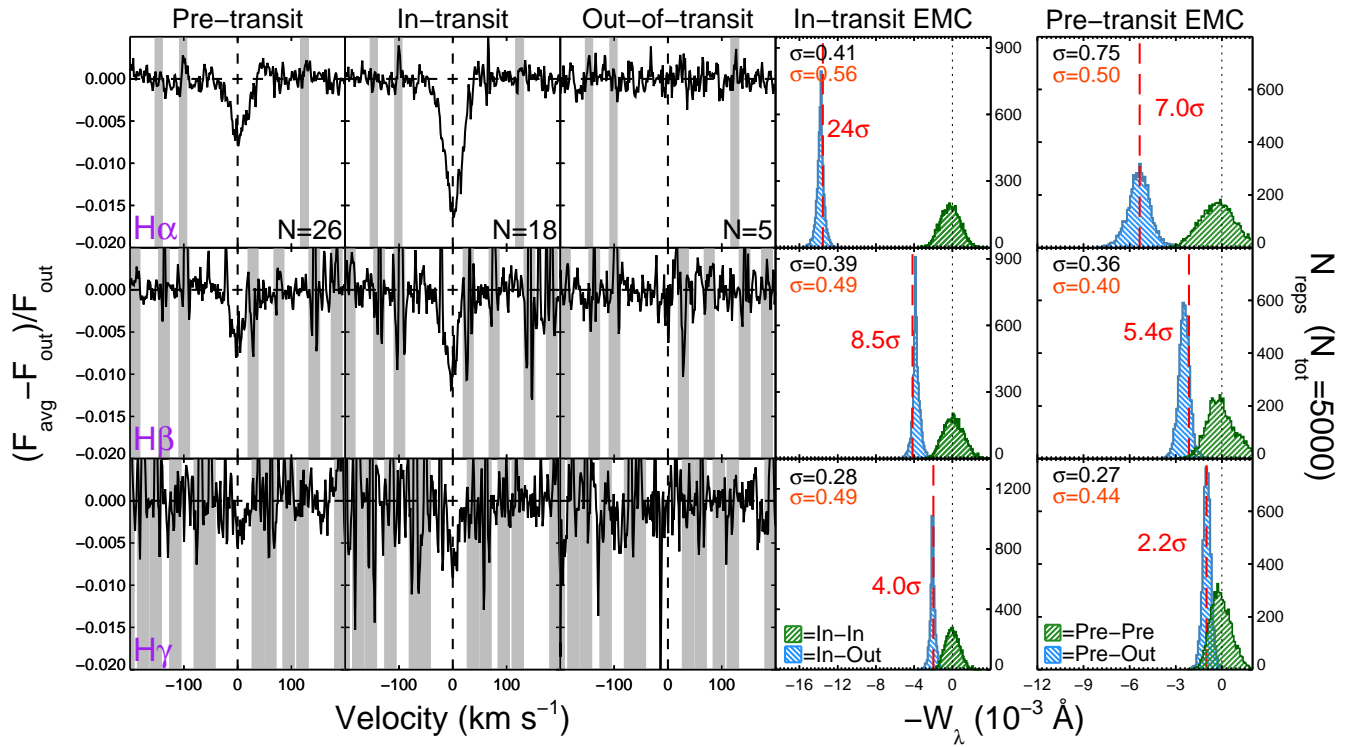} 
   \figcaption{The averaged transmission spectra for the pre-, in-, and post-transit spectra for H$\alpha$ (top row), H$\beta$ 
   (middle row), and H$\gamma$ (bottom row). The cores of identified stellar lines other than the lines of interest are 
   masked in gray. These points are not included in the calculation of the absorbed flux. The number of spectra used
   to create the average spectrum is listed in the first three panels of the top row. The fourth frame in each 
   row shows the in--transit empirical Monte Carlo (EMC) $W_\lambda$ distributions for each line. The master
   absorption measurement is marked with a vertical dashed red line. The uncertainty in W$_\lambda$ derived from
   the In-In EMC procedure is marked in black in the upper-left; the propagated flux uncertainty is labeled in orange. 
   The larger of the two is adopted for determining the detection significance. Each in-transit absorption measurement is 
   detected above the 3$\sigma$ level. The H$\alpha$ and H$\beta$ pre-transit measurements are detected at
   $>3\sigma$.\label{fig:tspecs_all}}
\end{figure*}

\begin{figure*}[h]
   \centering
   \includegraphics[scale=.68,clip,trim=8mm 55mm 5mm 40mm,angle=0]{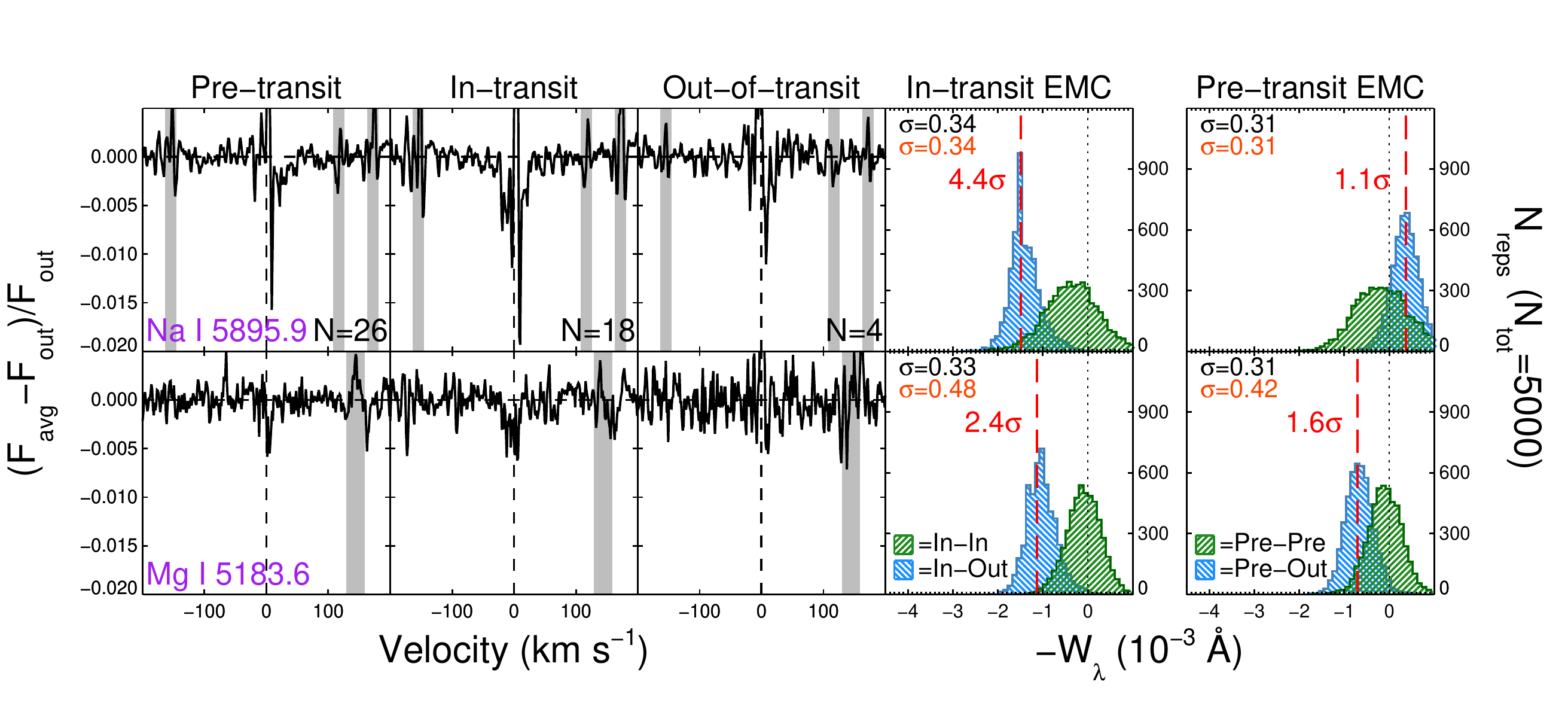} 
   \figcaption{The averaged transmission spectra for the pre-, in-, and post-transit spectra for \ion{Na}{1} 5895.9 \AA\ (top row) and
   \ion{Mg}{1} 5183.6 \AA\ (bottom row). All figure markings and colors
   are the same as \autoref{fig:tspecs_all}. The \ion{Na}{1} in-transit absorption is measured at 4.4$\sigma$ while
   the \ion{Mg}{1} absorption is marginal at 2.4$\sigma$. Neither line is detected in the pre-transit transmission spectra.\label{fig:tspecs_z}}
\end{figure*}

\subsection{Average transmission spectra}
\label{sec:ave_tspecs}

The average transmission spectra are shown in \autoref{fig:tspecs_all} for H$\alpha$ 6562.79 \AA\ (first row), 
H$\beta$ 4861.35 \AA\ (second row), and H$\gamma$ 4340.47 \AA\ (third row)
for the pre-transit (first column), in-transit (second column), and out-of-transit comparison
observations (third column). The number of spectra used to compute the average spectrum is shown in the
bottom right corner of each panel in the first row. The fourth column shows the empirical Monte Carlo (EMC) distributions,
which highlight the influence of systematic effects on the absorption measurements, of the
absorption for each line. The fifth column shows the EMC for the pre-transit signal. \autoref{fig:tspecs_z} shows
the same thing for the \ion{Na}{1} 5895.92 \AA\ and \ion{Mg}{1} 5183.60 \AA\ lines. Due to its location at the very edge of the order, we do not
extract and analyze the \ion{Na}{1} 5889 \AA\ line. 

\begin{figure*}[htbp]
   \centering
   \includegraphics[scale=.75,clip,trim=25mm 25mm 0mm 40mm,angle=0]{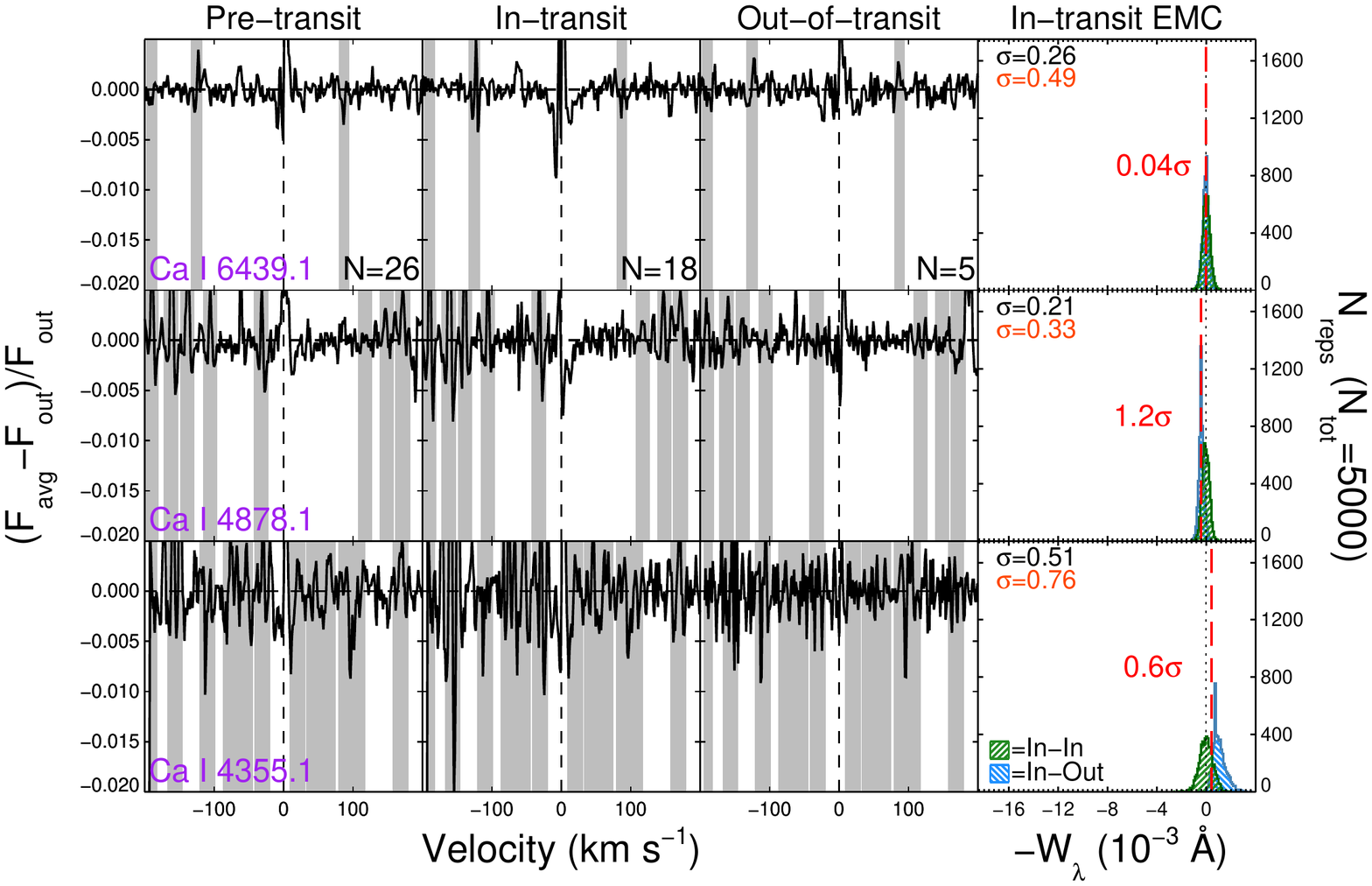} 
   \figcaption{The averaged transmission spectra for the pre-, in-, and post-transit spectra for the \ion{Ca}{1} control
   lines in each of the Balmer line orders. None of the control lines are detected at a significant level.The scale 
   of the EMC panels is the same as those in \autoref{fig:tspecs_all} to facilitate
   a direct comparison of the distributions. All of the colors and labels are the same as those in \autoref{fig:tspecs_all}.\label{fig:tspecs_ctrl}}
\end{figure*}

We perform two different EMC procedures: an In-In and In-Out procedure \citep[see][for similar applications]{redfield,jensen12,wyttenbach,cauley15}. 
The Pre-Pre and Pre-Out procedures are identical but for the pre-transit spectra.
The In-Out EMC compares randomly selected subsets of the in-transit spectra to the master comparison
spectrum. The number of random spectra varies from $N$=5 to $N$=18 and the process is repeated
for $N$=5000 iterations. The In-In procedure compares a random subset of in-transit spectra with
another random subset of in-transit spectra. The random subsets are chosen without replacement
so that spectra are never compared to themselves and the number of random spectra varies from
$N$=2 to $N$=16. The In-In distributions should be centered at zero absorption and should be
broader than the In-Out distributions due to the effect of comparing two different random subsets
instead of one random subset with a fixed comparison spectrum. We choose the standard deviation of the In-Out distribution
as the 1$\sigma$ uncertainty associated with the EMC procedure since this is directly related to
the variation in the absorption measurement.
 
As expected, the In-In distributions (green) are centered near zero absorption.  
The 1$\sigma$ values from the In-Out EMC procedure are shown in black. We also calculate the propagated
flux uncertainty associated with the master absorption measurement (vertical red dashed line). 
The flux uncertainty is the standard deviation of two 50 km s$^{-1}$ wide portions of the transmission spectrum 
from $-250$-$-200$ km s$^{-1}$ and $+200$-$+250$ km s$^{-1}$ weighted by the square root
of the normalized observed spectrum in order to account for the larger uncertainties in the line cores.
This value is shown in orange. The larger of the EMC and flux uncertainty measurements is used
for determining the significance of the absorption. The significance level $n\sigma$ is shown in red
next to the master absorption measurement.

The absorption is the negative of the equivalent width of the transmission spectrum integrated
from $-200$ km s$^{-1}$ to $+200$ km s$^{-1}$. More formally,

\begin{equation}\label{eq:wlambda}
W_\lambda = \sum\limits_{v=-200}^{+200} \left(1-\frac{F_v}{F_v^{out}} \right) \Delta\lambda_v
\end{equation}

\noindent where $F_v$ is the flux in the spectrum of interest at velocity $v$, $F_v^{out}$ is the flux in the
comparison spectrum at velocity $v$, and $\Delta\lambda_v$ is the wavelength difference at velocity
$v$. The gray regions in \autoref{fig:tspecs_all} indicate the cores of identified stellar absorption lines other than
the line of interest. These regions are ignored when calculating $W_\lambda$. The master absorption
detections are presented in \autoref{tab:tab1}. 

The in-transit and pre-transit absorption for H$\alpha$ and H$\beta$ is detected at $>3\sigma$, confirming the 
detections made in \citetalias{cauley15} and the H$\alpha$ detection from \citet{jensen12}. The level of in-transit
H$\alpha$ absorption is almost identical to that from \citet{jensen12} ($W_{H\alpha}$=13.7$\times$10$^{-3}$ \AA)
and 2.3$\times$ the value measured in \citetalias{cauley15} ($W_{H\alpha}$=5.9$\times$10$^{-3}$ \AA). The in-transit
H$\gamma$ absorption is detected at 4.0$\sigma$; the measurement is marginal (2.2$\sigma$) for the pre-transit
absorption. 

The in-transit \ion{Na}{1} absorption is detected at 4.4$\sigma$ while the \ion{Mg}{1} absorption is marginal
at 2.4$\sigma$. There is little evidence for pre-transit absorption in either line. In order to compare
our \ion{Na}{1} 5896 \AA\ in-transit absorption value to previously reported high resolution \ion{Na}{1} absorption, we have 
applied \autoref{eq:wlambda} to the \ion{Na}{1} 5896 \AA\ transmission spectra presented in \citet{redfield}, 
\citet{jensen11}, and \citet{wyttenbach}. In the case of \citet{wyttenbach}, we have used their reported Gaussian fit parameters 
for the 5896 \AA\ line in order to approximate the absorption. For \citet{jensen11} we find $W_{NaI}$=2.85$\pm$0.85$\times10^{-3}$ \AA; 
for \citet{redfield}, $W_{NaI}$=3.59$\pm$1.61$\times$10$^{-3}$ \AA; for \citet{wyttenbach}, $W_{NaI}$=2.22$\times$10$^{-3}$ \AA.
Our reported value of $W_{NaI}=$1.50$\pm$0.34$\times$10$^{-3}$ \AA\ is consistently lower than the previous measurements. 

Although the \ion{Mg}{1} 5184 \AA\ in-transit absorption is marginal, the measurement is suggestive of a real signal. 
The \ion{Mg}{1} UV transition at 2853 \AA\ has been detected previously for HD 209458 b by \citet{vidal13}, providing
evidence that \ion{Mg}{1} is present at detectable levels in hot Jupiter atmospheres. The 
HD 209458 b signal is much stronger (7.5\%) than that reported here. 
Although the \ion{Mg}{1} 5184 \AA\ line arises from an excited \ion{Mg}{1} state and thus the 
line strength is not favorable compared to the UV resonance line, future investigations should consider examining the 
\ion{Mg}{1} 5184 \AA\ line due to its ease of observability compared to the UV \ion{Mg}{1} lines \citep[see ][]{bourrier15}
and to confirm or reject the measurement presented here.

%\capstartfalse           
\begin{deluxetable}{lccccc}
%\rotate
%\linespacing{1}
\tablecaption{Master absorption line detections \label{tab:tab1}}
%\tablewidth{0pt}
\tablehead{\colhead{}&\multicolumn{2}{c}{Pre-transit}&\colhead{}&\multicolumn{2}{c}{In-transit}\\
\cline{2-3}\cline{5-6}\\
\colhead{}&\colhead{W$_\lambda$}&\colhead{$\sigma$}&\colhead{}&\colhead{W$_\lambda$}&\colhead{$\sigma$}\\
\colhead{Spectral line}&\colhead{(10$^{-3}$ \AA)}&\colhead{(10$^{-3}$ \AA)}&\colhead{}&\colhead{(10$^{-3}$ \AA)}&\colhead{(10$^{-3}$ \AA)}\\
\colhead{(1)}&\colhead{(2)}&\colhead{(3)}&\colhead{}&\colhead{(4)}&\colhead{(5)}}
\tabletypesize{\scriptsize}
\startdata
H$\alpha$ & 5.25 & 0.75 & & 13.44 & 0.56 \\
H$\beta$ & 2.16 & 0.40 & & 4.17 & 0.49\\
H$\gamma$ & 0.97 & 0.44 & & 1.96 & 0.49 \\
\ion{Na}{1} 5896 & 0.34 & 0.31 & & 1.50 & 0.34 \\
\ion{Mg}{1} 5184 & 0.67 & 0.42 & & 1.15 & 0.48 \\
 & & & &  &  \\
\ion{Ca}{1} 6439.1 & \nodata & \nodata & & 0.02 & 0.49 \\
\ion{Ca}{1} 4878.1 & \nodata & \nodata & & 0.40 & 0.33 \\
\ion{Ca}{1} 4355.1 & \nodata & \nodata & & $-$0.46 & 0.76 \\
\enddata
\end{deluxetable}
%\capstarttrue

\subsection{\ion{Ca}{1} control lines}
\label{sec:cai_control}

\autoref{fig:tspecs_ctrl} shows a set of \ion{Ca}{1} control lines that are used to verify the transmission
spectrum analysis. Calcium is expected to condense out of hot Jupiter atmospheres and so should
not be present in any significant quantity \citep{lodders}. Each of the \ion{Ca}{1} lines is located in the
same order as the Balmer line closest to it in wavelength. \autoref{fig:tspecs_ctrl} is identical to
\autoref{fig:tspecs_all} except we do not perform a pre-transit EMC for the control lines. Using the
flux errors as the 1$\sigma$ uncertainties, none of the control lines are measured at a significant
level (see \autoref{tab:tab1}). The non-detections in the control lines provide reassurance that the
measurement algorithms are being properly applied and not resulting in spurious absorption
signatures. 

\subsection{Absorption time series}
\label{sec:indtranspec}

\begin{figure*}[htbp]
   \centering
   \includegraphics[scale=.80,clip,trim=25mm 25mm 5mm 20mm,angle=0]{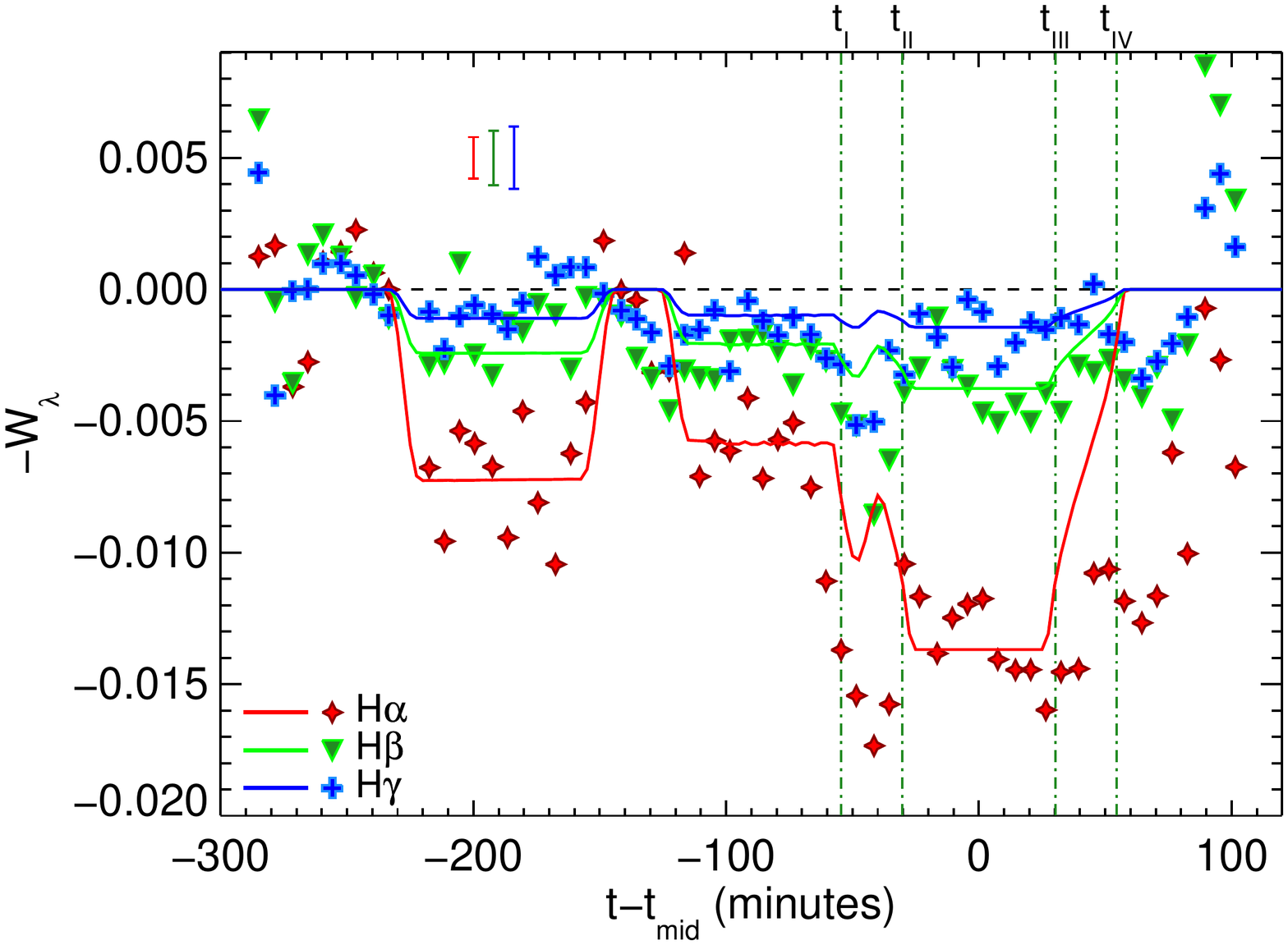} 
   \figcaption{Absorption time series calculated for each of the transmission spectra shown in \autoref{fig:tspecs_halpha},
   \autoref{fig:tspecs_hbeta}, and \autoref{fig:tspecs_hgamma}. Values of W$_\lambda$ are calculated using
   \autoref{eq:wlambda}. The average 1$\sigma$ uncertainties in W$_\lambda$ for each line
   are marked with the solid bars in the upper-left. Optical transit contact points are marked with
   vertical green dashed-dotted lines. The solid lines are model $W_\lambda$ values (see \autoref{sec:model}). Note 
   the two distinct pre-transit dips in W$_{H\alpha}$ between $-220$ minutes
   and $-155$ minutes and then again between $-110$ and $-60$ minutes. The in-transit absorption strength is $\sim$2
   times stronger than observed in the 2013 transit and there appears to be sustained absorption for $\sim$30 minutes
   post-transit.\label{fig:mabs_all}}
\end{figure*}

\begin{figure*}[htbp]
   \centering
   \includegraphics[scale=.80,clip,trim=25mm 25mm 5mm 20mm,angle=0]{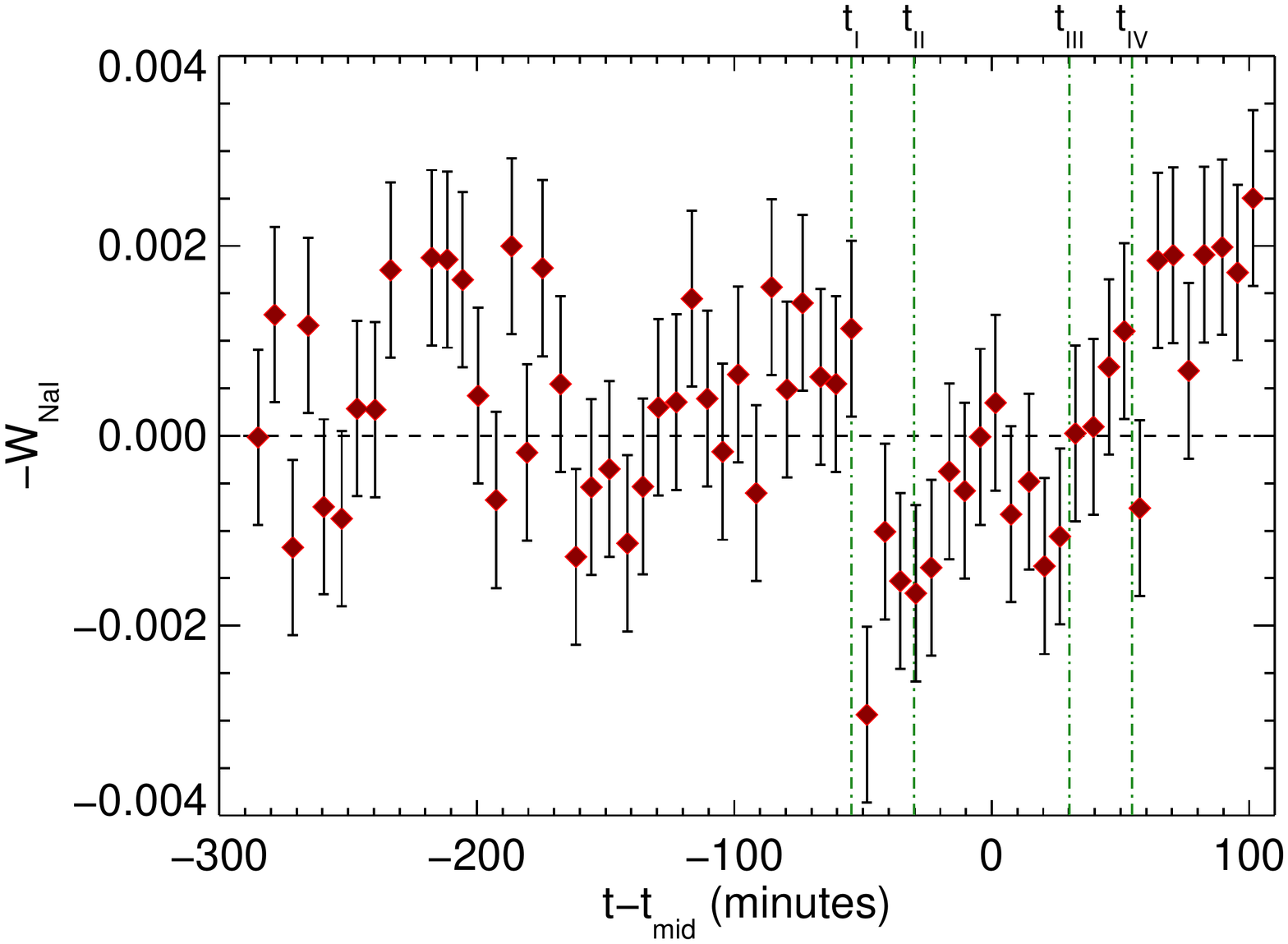} 
   \figcaption{Absorption time series calculated for each of the \ion{Na}{1} transmission spectra shown in \autoref{fig:tspecs_nai}.\label{fig:mabs_nai}}
\end{figure*}

\begin{figure*}[htbp]
   \centering
   \includegraphics[scale=.80,clip,trim=25mm 25mm 5mm 20mm,angle=0]{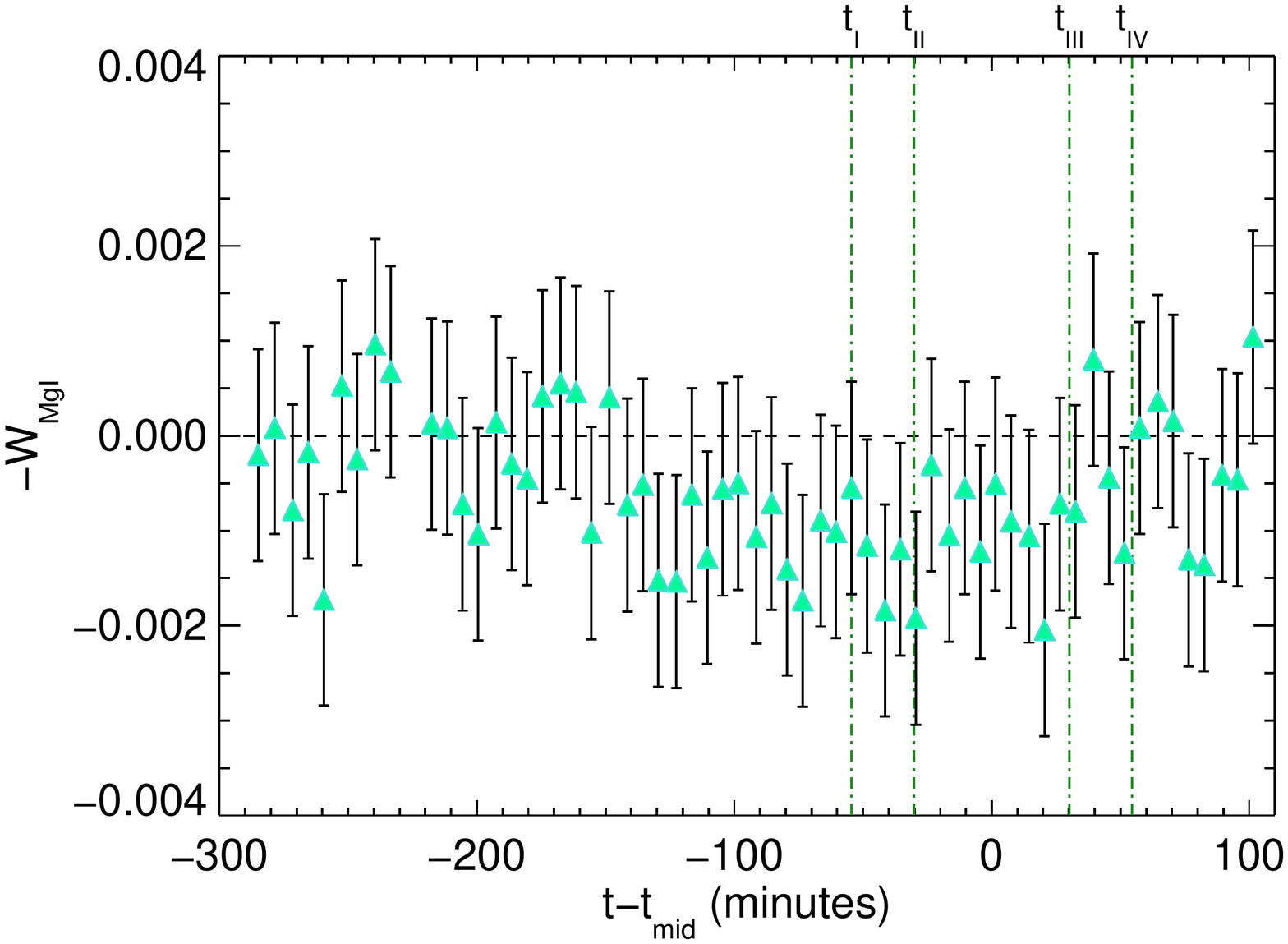} 
   \figcaption{Absorption time series calculated for each of the \ion{Mg}{1} transmission spectra shown in \autoref{fig:tspecs_mgi}.\label{fig:mabs_mgi}}
\end{figure*}

Transmission spectra for individual exposures are shown in \autoref{sec:tspecs_app}. 
The values of W$_\lambda$ for the individual spectra are shown as a time series in \autoref{fig:mabs_all}, \autoref{fig:mabs_nai},
and \autoref{fig:mabs_mgi}. Each of the values is calculated using \autoref{eq:wlambda}. For all measurements, we choose to use the more 
conservative average flux uncertainties for the individual points rather than the standard deviation of the EMC
procedure. In \autoref{fig:mabs_all}, the mean uncertainty for the 
individual measurements is shown with the solid colored bars in the upper left. Individual $1\sigma$ uncertainties 
are shown for \ion{Na}{1} and \ion{Mg}{1} in \autoref{fig:mabs_nai} and \autoref{fig:mabs_mgi}.

The pre-transit Balmer line signature in \autoref{fig:mabs_all} begins abruptly at $-217$ minutes, which corresponds to a linear distance of
$\sim$17 R$_p$ to the stellar limb, and lasts for $\sim$70 minutes until it abruptly disappears. It then returns to a 
similar level after $\sim$40 minutes and remains relatively constant until immediately before first contact. The in-transit absorption appears to begin 
immediately before first contact and increases between first and second contact. The abrupt changes in the level
of in-transit absorption, e.g., at $-30$, 5, and 40 minutes, may be due to the planet transiting the inhomogeneous
stellar surface. \added{The first event begins at $\sim-$70 minutes, or $\sim$15 minutes before $t_I$. This suggests
that the atmosphere is occulting an extended prominence above the stellar disk. This event ends at $t_{II}$, immediately after
the planetary disk fully occults the star. The second event, from $\sim+$5 to $+40$ minutes, has a duration similar
to the first event. The duration of the events correspond to stellar features with a linear extent of $\sim$2 $R_p$. These
abrupt changes in the absorbed flux could also be due to changes in the stellar activity level rather than transits
of static features. While this cannot be definitively ruled out, we believe this is less likely due to the very similar duration 
of both events.} The effect of active regions and changes in the stellar activity level on the line absorption 
will be discussed more fully in \autoref{sec:activityimpact}.

An interesting feature of the 2015 transit is that the absorption appears to persist immediately post-transit for $\sim$30 minutes. The subsequent decrease
in absorption at $\sim$90 minutes, and appearance of emission in the line cores of H$\beta$ and H$\gamma$, may be due to a change
in the stellar activity level, e.g., a mild flaring event. After the sharp increase, the final two points in the time series
decrease from the maximum. The linear extent of the feature causing the post-transit absorption is $>$3 $R_p$ 
if we take the beginning of the small flaring event to be the end of the post-transit absorption. The depth of
the measured post-transit absorption indicates a disk coverage fraction larger than the nominal extended atmosphere
we use to model the in-transit absorption in \autoref{sec:model} since the opaque planet is no longer blocking the
stellar disk. The post-transit absorption may be the base of the extended evaporative flows previously detected
around HD 189733 b in neutral hydrogen \citep{desetangs,desetangs12,bourrier13}.

Only one individual $W_\lambda$ value at $t-t_{mid}$=$-48$ minutes, is detected at $>$3$\sigma$ for \ion{Na}{1}. 
None of the individual absorption values for \ion{Mg}{1} is detected at $>2\sigma$. There is a notable decrease, however,
in the \ion{Na}{1} in-transit points compared to the pre- and post-transit values and the \ion{Mg}{1} values are
consistently below zero beginning at $\sim$$-255$ minutes. The structure of the in-transit \ion{Na}{1} absorption
shows a similar shape to the center-to-limb variations (CLVs) described by \citet{czesla15}. CLVs may be
important in determining the precise level of atmospheric \ion{Na}{1} absorption and should be taken
into account for a detailed analysis of the in-transit absorption. Such an analysis is beyond the scope of
this paper. The consistently high post-transit values of $W_{NaI}$ in \autoref{fig:mabs_nai} appear to
be due to the large core residuals and not to any real emission features.

\subsection{H$\alpha$ line centroid velocities}
\label{sec:havels}

The signal-to-noise in the individual H$\alpha$ transmission spectra is high enough to consider the overall velocity
shift of the absorption. Measurements of the line velocity can be useful in understanding the mass motions of the
absorbing material. In particular, measurements of in-transit line velocities can provide information about
the planetary rotation and atmospheric dynamics \citep[e.g.,][]{showman02,menou10,rauscher14,snellen14,louden,brogi15}.
Here we will focus on the pre-transit line velocities and a detailed analysis of the in-transit velocities
will be presented in a future paper.

\begin{figure*}[htbp]
   \centering
   \includegraphics[scale=.80,clip,trim=30mm 25mm 5mm 20mm,angle=0]{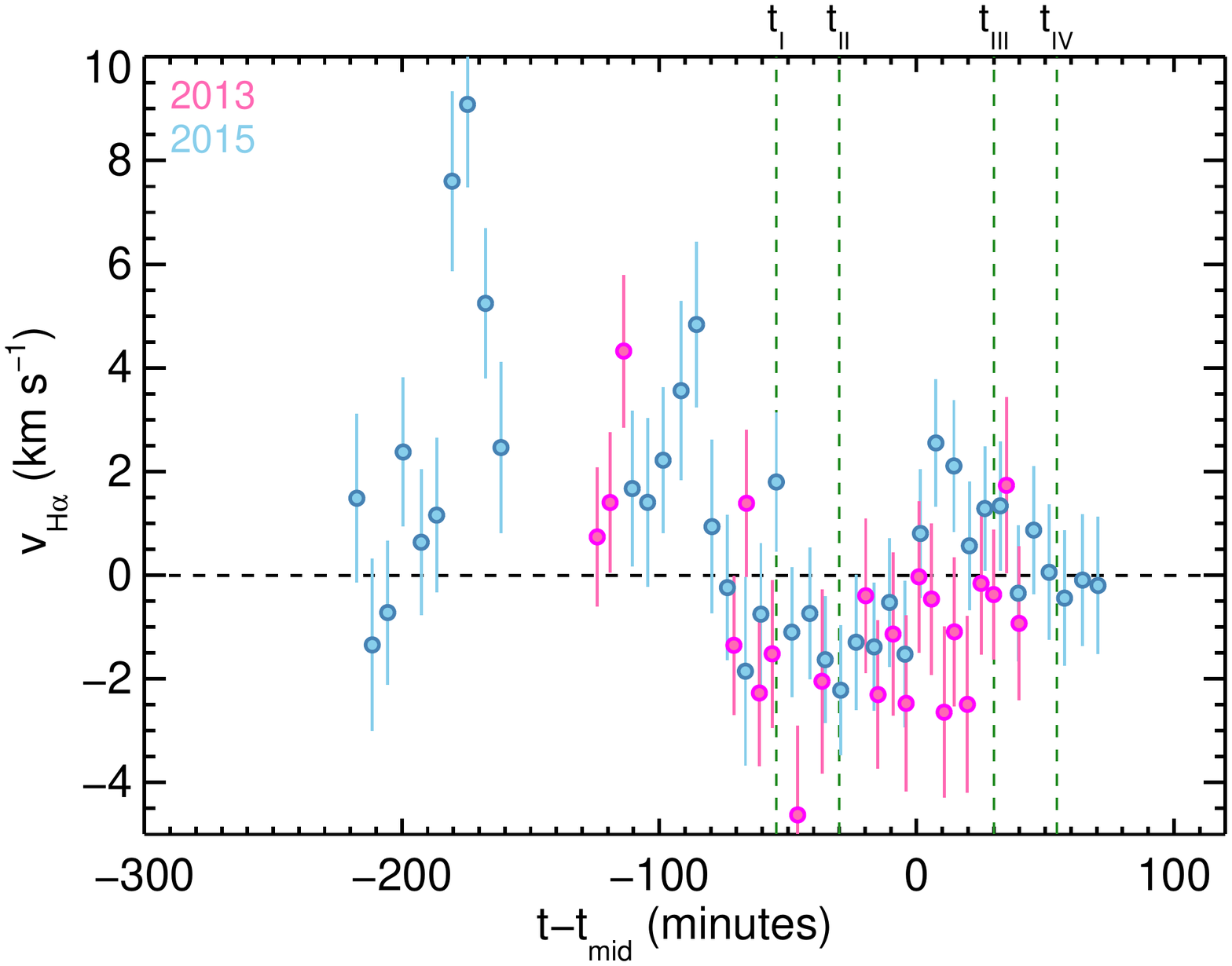} 
   \figcaption{Line centroid velocities calculated for H$\alpha$ transmission spectra that display 2$\sigma$ absorption in \autoref{fig:tspecs_halpha}.
   Values calculated for the 2013 data from \citetalias{cauley15} are also shown. The green vertical dashed lines show the transit contact points. 
   The full range of $t-t_{mid}$ is shown for easy comparison with \autoref{fig:mabs_all}. The pre-transit lines show mostly positive velocities,
   indicating material moving towards the star. An analysis of the in-transit velocities will be presented in a future paper.\label{fig:halinevels}}
\end{figure*}

The velocity is measured for each H$\alpha$ transmission spectrum which has a W$_\lambda$ value significant at
the 2$\sigma$ level. We use a flux-weighted average from $-40$ to $+40$ km s$^{-1}$ to measure the velocity of the absorption:

\begin{equation}\label{eq:vhalpha}
v_{H\alpha} = \frac{\sum\limits_{v=-40}^{+40} v \left(1-f_{v}\right)^2}{\sum\limits_{v=-40}^{+40} \left(1-f_{v}\right)^2}
\end{equation}

\noindent where $(1-f_{v})^2$ is the square of the transmission spectrum flux at velocity $v$. This
ensures that deeper portions of the line are heavily weighted. We also add back in the velocity corrections from \autoref{fig:halphavcorrs}.
We note that these corrections are small ($\sim$-1.0 to 1.0 km s$^{-1}$) and do not significantly affect any of the conclusions
based on the line velocities discussed here. The uncertainty in each velocity is the standard deviation of the 
distribution derived from calculating the line velocity for each of the $N=255$ combinations ($N=511$ for the 2013 data set) used
to produce the individual spectra in \autoref{fig:tspecs_halpha}. This is essentially the EMC process introduced in \autoref{sec:indtranspec}.
We add an extra 1.0 km s$^{-1}$ to each uncertainty to approximate the error in the original velocity correction.

The H$\alpha$ velocities for both the 2013 and 2015 data sets are shown in \autoref{fig:halinevels}. Most of the 2015 pre-transit
velocities are red-shifted, indicating material moving away from the observer. However, immediately before the transit the
velocities are grouped relatively close to zero with a sharp transition down from $\sim+$4 km s$^{-1}$ between $t-t_{mid}=-85$ and 
$-79$ minutes. There is no obvious explanation for this abrupt change in the line velocity. Also note the consistent blue-shift
of the 2013 in-transit points compared to the symmetric in-transit velocities of the 2015 data. We will revisit the pre-transit
velocities during \autoref{sec:model} in the context of the clumpy accretion stream model.

\section{The residual \ion{Ca}{2} H core flux}
\label{sec:calcium2}

The \ion{Ca}{2} H and K lines are frequently used as stellar activity indicators \citep[e.g.,][]{duncan,wright04,isaacson,gomes14} and are standard
diagnostics of solar flare energetics and dynamics \citep[e.g.,][]{johnskrull97,garcia05,lalitha}. H$\alpha$, which is in emission
in stellar active regions, is also used as an activity measure \citep{meunier09,gomes14,kuridze} and correlates directly with \ion{Ca}{2} H and K
across solar cycles \citep{livingston07}. Much variation exists, however, in the H$\alpha$-\ion{Ca}{2} H and K correlation
for other stars \citep[e.g.,][]{cincunegui}. Surface H$\alpha$ activity can therefore mimic an absorption
signature: if the comparison spectra capture the star in a more active state compared to the spectra of interest (e.g., the 
pre- or in-transit spectra), the filling in of the line cores in the active spectra will result in ``absorption'' in the transmission spectrum
of a non-active spectrum. Thus it is critical to understand the stellar activity level as a function of time in order
to attempt to separate any activity contribution to the absorption signature. 

While measurements of \ion{Ca}{2} and H$\alpha$ 
core flux have been used to study the long term (days to years) influence of hot planets on their host stars 
\citep[e.g.,][]{shkolnik05,shkolnik08,fares10,scandariato}, there has been little effort to characterize the very short term variations
(minutes to hours) in the line cores of stars other than the Sun. Ideally, short-cadence observations of the star would be carried out at
multiple phases of a single planetary orbit in order to establish a baseline of the short-term stellar activity level and
characterize typical changes in the line core flux. This baseline could be used to separate the purely stellar component
from any variations caused by circumstellar material during the phase of interest. 

H$\alpha$ and \ion{Ca}{2} H and K emission is spatially coincident in active regions on the Sun
\citep[e.g., see][]{meunier09,kuridze}, although the two line fluxes do not necessarily correlate
across the entire solar disk. Filaments, which are cooler than plages and can absorb active region
H$\alpha$ emission, can affect the level of observed H$\alpha$ flux and weaken the correlation
between H$\alpha$ and \ion{Ca}{2} H and K \citep{meunier09}.  Short cadence observations of flaring
regions on the Sun show a direct correlation between \ion{Ca}{2} and H$\alpha$ excess emission
equivalent widths \citep[e.g.,][]{johnskrull97,garcia05}. A similar correspondence is seen in dMe
flares. In the case of dMe objects, the peak \ion{Ca}{2} K flux occurs $\sim$30 minutes after the
peak in the Balmer line fluxes and variations of $\sim$5\% are common on 10-minute timescales
\citep{kowalski13}. Although HD 189733 is not as active as a typical dMe star, it is an active
K-dwarf and so it is likely that it would exhibit low-level chromospheric variability on
short timescales. To the best of our knowledge, no continuous short-cadence observations of HD
189733's H$\alpha$ and the \ion{Ca}{2} H and K lines, similar to the observations presented here and
in \citetalias{cauley15}, have been published. Thus it is difficult, outside of the transits
presented in this work, to quantify the stellar contribution to variations in the line fluxes on
short timescales. 

\begin{figure}[htbp]
   \centering
   \includegraphics[scale=.65,clip,trim=65mm 35mm 5mm 70mm,angle=0]{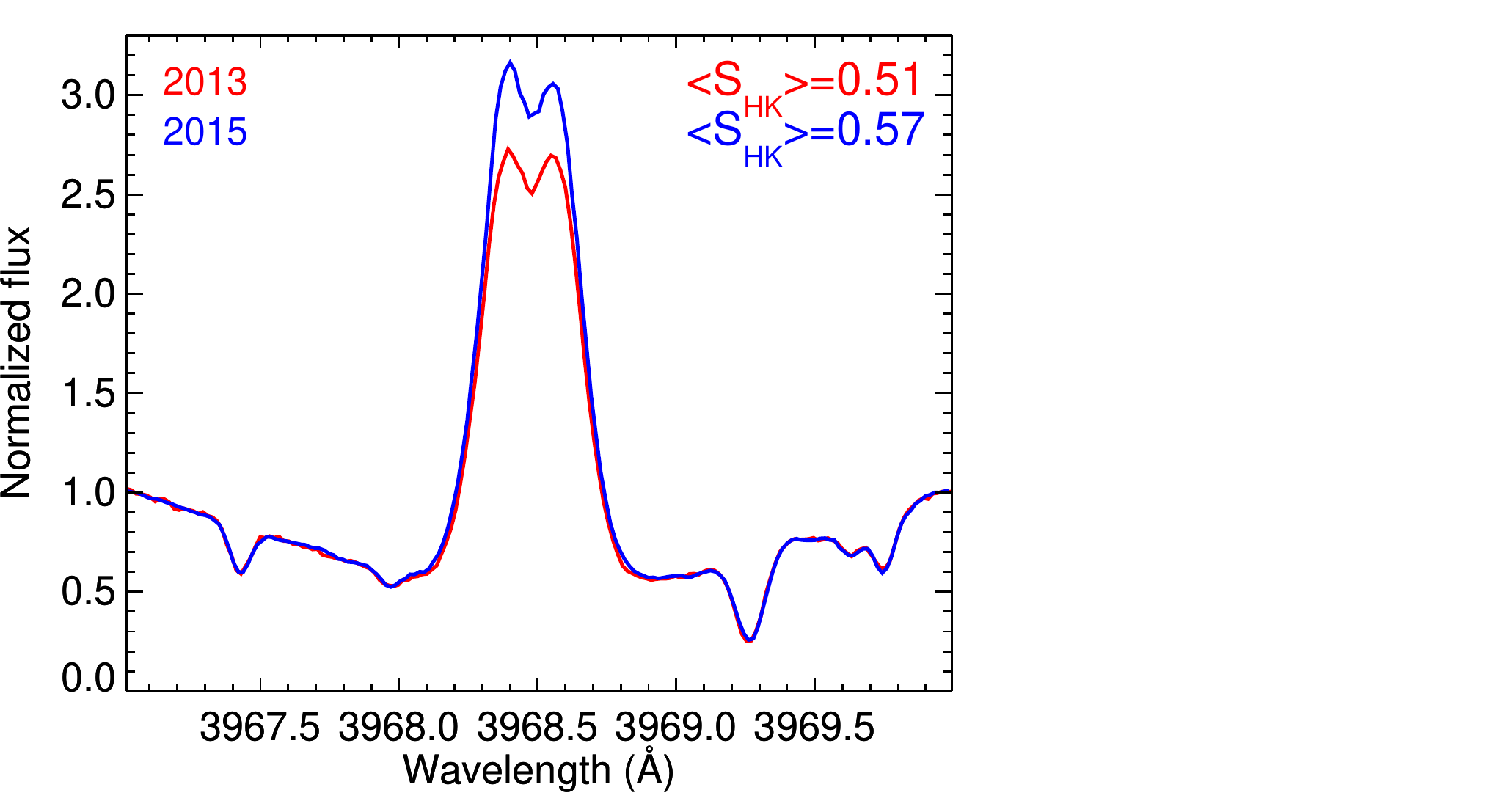} 
   \figcaption{Comparison of the normalized \ion{Ca}{2} H line cores of the comparison spectrum for the 2013 transit (red) 
   and 2015 transit (blue). The 2015 line core has $\sim$12\% more flux than the 2013 line core, indicating that the
   star is in a more active state during the 2015 transit.\label{fig:caii_comp}}
\end{figure}

In \citetalias{cauley15} we used the time series structure of the \ion{Ca}{2} Mt. Wilson S$_{HK}$
index to argue that the pre-transit signal was not caused by varying stellar activity and was
instead due to material occulting the star. This argument was based on the lack of correlation
between the S$_{HK}$ index and W$_{H\alpha}$. Furthermore, the mean S$_{HK}$ value of the comparison
spectra was almost identical to that of the pre-transit spectra, suggesting a similar activity level
pre- and post-transit. However, the S$_{HK}$ index compares the core flux with a wide continuum
window and, as a result, is subject to variations in the continuum flux. Small variations in the
continuum window flux, which are not necessarily related to the same physical processes causing flux
changes in the line cores, can thus mimic changes in the stellar activity level.

In order to mitigate any contribution to the activity index by the continuum windows, we have reanalyzed the
2013 \ion{Ca}{2} lines using the residual core flux method of \citet{shkolnik05}. We have also applied this
analysis to the 2015 \ion{Ca}{2} lines. To be clear, the S$_{HK}$ index is well suited for measuring average
activity levels across different epochs and, indeed, we present the average S$_{HK}$ values for each night
in \autoref{fig:caii_comp}. However, a more precise measurement of variations in the stellar activity level,
is better achieved by measuring only the core flux. 

The residual profiles are constructed by normalizing the line core to 0.1 \AA\ wide spectral regions
centered at 3967.05 \AA\ and 3969.95 \AA\ and subtracting the mean profile of the same comparison
spectra used to generate the average Balmer line transmission spectra. Individual values of
W$_{CaH}$ are calculated by summing over the residual spectrum from 3967 \AA\ to 3970 \AA\ . The
mean \ion{Ca}{2} H profiles for the comparison spectra from each night are shown in
\autoref{fig:caii_comp}. The \ion{Ca}{2} H residual profiles are shown for the 2013 transit in
\autoref{fig:caiires_0704} and for the 2015 transit in \autoref{fig:caiires_0804_1} and
\autoref{fig:caiires_0804_2}. We exclude the \ion{Ca}{2} K profiles due to lower
signal-to-noise.\footnote{The \ion{Ca}{2} H and K lines are located near the edges of separate
orders in our spectrograph setup.} We note that the residual profiles are not identical to the
transmission spectra, since they are not normalized by the comparison spectrum, and so values of
W$_{H\alpha}$ and W$_{CaH}$ cannot be directly compared.

\begin{figure*}[htbp]
   \centering
   \includegraphics[scale=.80,clip,trim=25mm 25mm 5mm 20mm,angle=0]{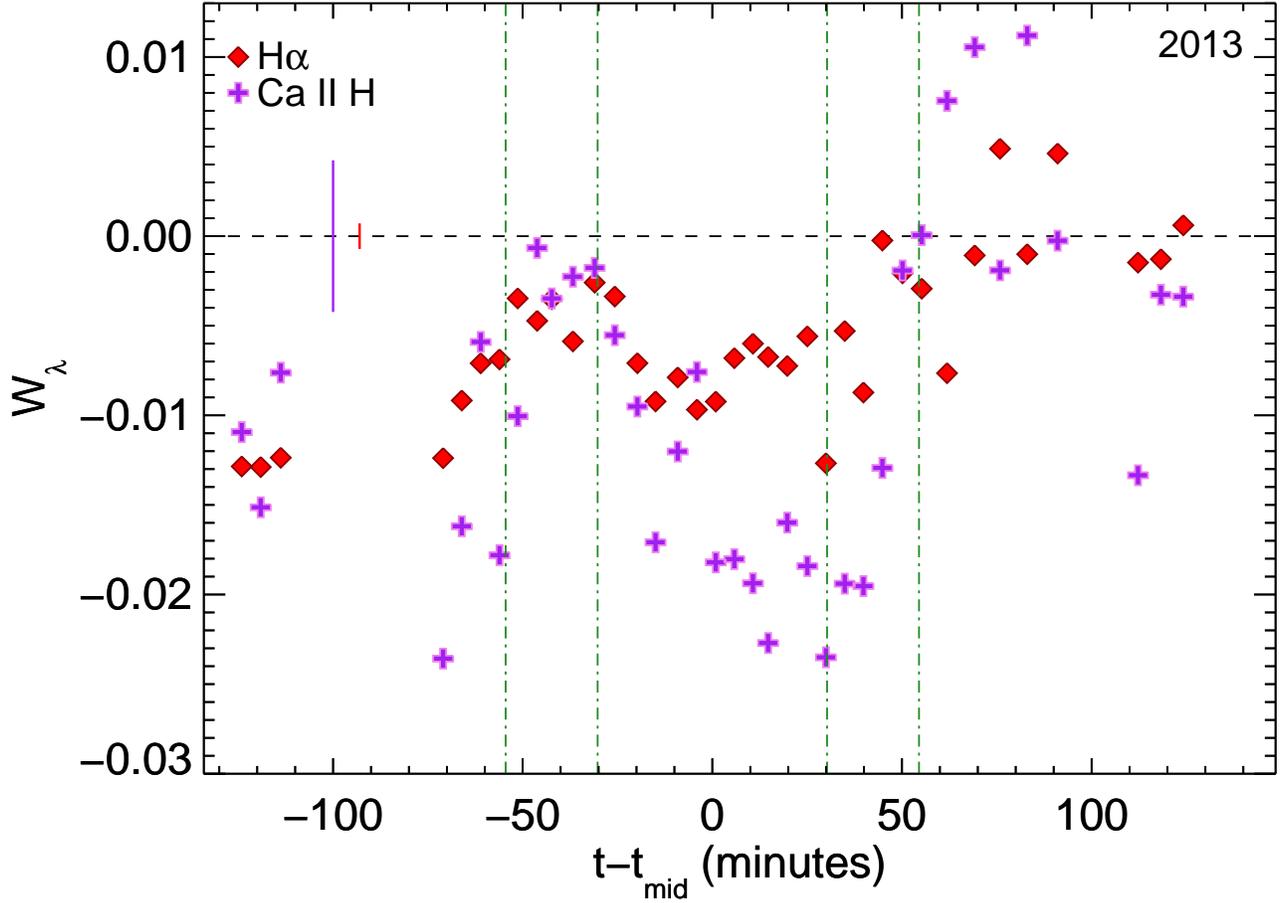} 
   \figcaption{\ion{Ca}{2} H residual core flux time series for the 2013 transit. The W$_{H\alpha}$ values are over plotted
   for comparison. Typical 1$\sigma$ uncertainties are shown with solid colored lines in the upper left. The core flux 
   alone shows different behavior than the S$_{HK}$ index used in \citetalias{cauley15}.
   Notably, the core flux appears to trace the H$\alpha$ absorption, suggesting that some of the signal in H$\alpha$
   attributed to absorption may be caused by varying stellar activity. In addition, the residual core flux of the
   comparison spectra (the nine post-transit spectra) is higher than the pre-transit spectra. However, the correlation is weak (see \autoref{fig:caii_corr}
   and \autoref{fig:rhos_dist}) so it is difficult to quantify how much of the signal is purely due to stellar activity.\label{fig:caiihatime_0704}}
\end{figure*}

\begin{figure*}[htbp]
   \centering
   \includegraphics[scale=.80,clip,trim=25mm 25mm 5mm 20mm,angle=0]{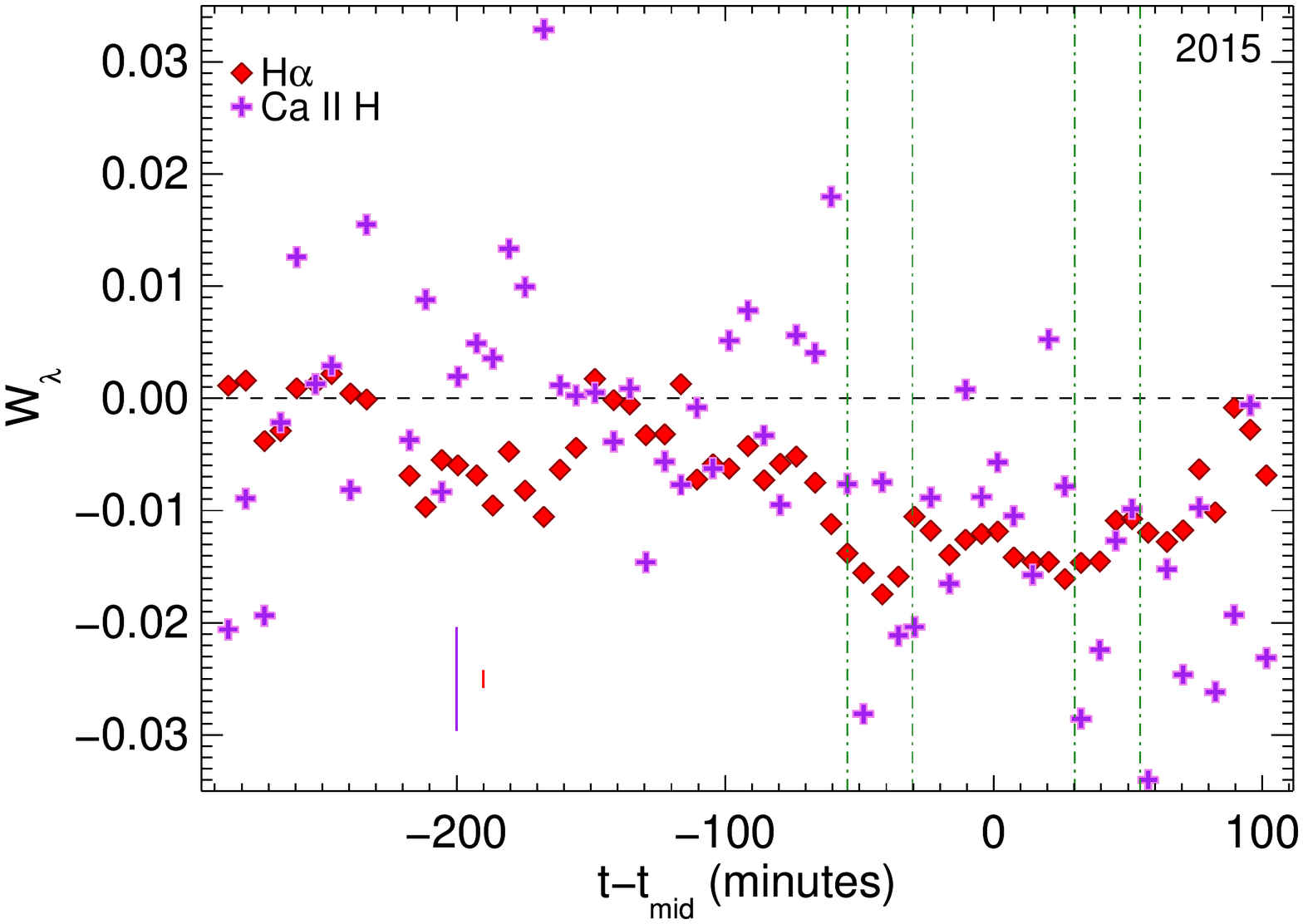} 
   \figcaption{The same as \autoref{fig:caiihatime_0704} except for the 2015 transit. Note the larger plot range for
   W$_\lambda$. The scatter in W$_{CaH}$ is larger than for the 2013 transit and there is little correlation between
   W$_{CaH}$ and W$_{H\alpha}$ (see \autoref{fig:caii_corr}), although the mean in-transit value of W$_{CaH}$
   is lower than the mean pre-transit value. Note the low values of W$_{CaH}$ for the first three comparison spectra
   near $\sim-280$ minutes.\label{fig:caiihatime_0804}}
\end{figure*}

\begin{figure}[htbp]
   \centering
   \includegraphics[scale=.58,clip,trim=35mm 15mm 5mm 10mm,angle=0]{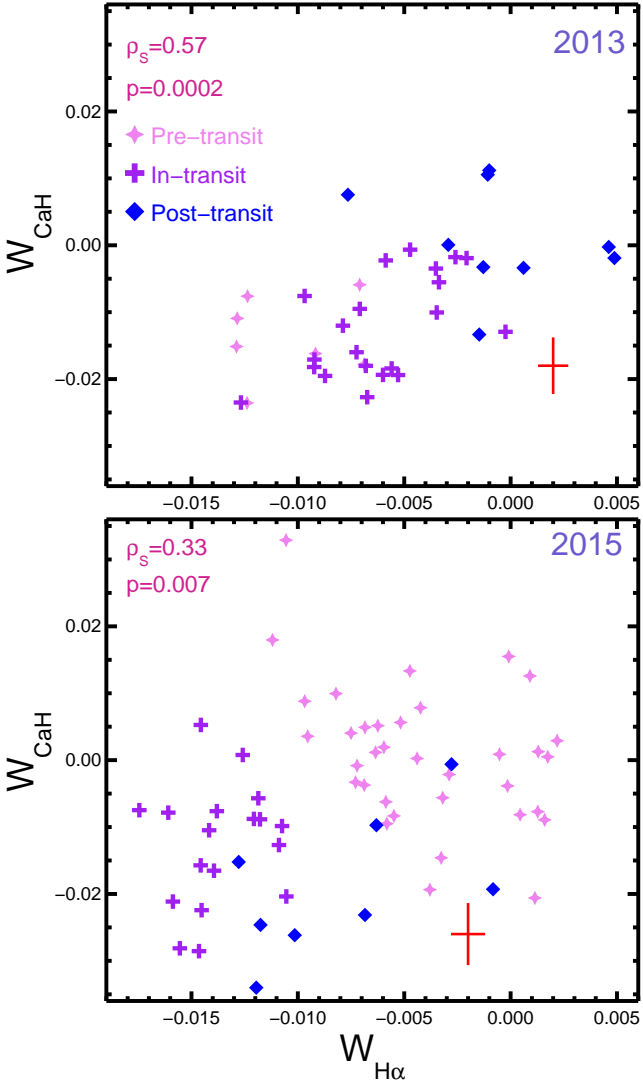} 
   \figcaption{W$_\lambda$ for \ion{Ca}{2} H versus H$\alpha$ for the 2013 and 2015 transits. Note the larger plot range for W$_{CaH}$
   in the bottom panel. There is a modest correlation
   in the 2013 data but it is driven almost entirely by the in-transit points ($\rho_s$=0.52, $p$=0.016; green crosses). There is a weak but significant
   correlation in the 2015 data. Typical 1$\sigma$ uncertainties in W$_\lambda$ are shown with the red solid lines.\label{fig:caii_corr}}
\end{figure}

\autoref{fig:caiihatime_0704} and \autoref{fig:caiihatime_0804} show the W$_{CaH}$ timeseries for
the 2013 and 2015 transits, respectively. The values of W$_{H\alpha}$ from each transit are shown
for comparison. Representative 1$\sigma$ uncertainties are shown with the solid colored bars. The
residual core flux in \autoref{fig:caiihatime_0704} shows different structure compared to the
S$_{HK}$ index from \citetalias{cauley15} \citepalias[see Figure 4 from ][]{cauley15}. Notably, the
pre-transit W$_{CaH}$ values appear to trace the pre-transit W$_{H\alpha}$ values and the mean
pre-transit W$_{CaH}$ value is lower than the mean post-transit value, suggesting similar levels of
stellar activity both pre- and post-transit. This contradicts the behavior of the S$_{HK}$ index
from \citetalias{cauley15} which showed very similar pre- and post-transit values. As described
above, this must be due to small changes in the continuum window flux used in the index calculation
since the core flux remains the same in both measurements. The 2013 \wca\ values are plotted against
the \wha\ values in the top panel of \autoref{fig:caii_corr}. The Spearman's $\rho$ value, $\rho_S$,
and the corresponding $p$-value are given in the upper left of the panel. There is a moderate
correlation between \wca\ and \wha, although we note that this is mainly driven by the in-transit
values (green crosses) which have $\rho_S$=0.51 ($p$=0.016).  The \ion{Ca}{2} H residual values for
the 2015 transit are shown in \autoref{fig:caiihatime_0804}. The 2015 W$_{CaH}$ and W$_{H\alpha}$ do
not show any obvious correspondence, although the in-transit \wca\ values are lower on average than
the pre-transit values. There is a weak but significant correlation present between the 2015 values
of \wca\ and \wha.

\begin{figure}[htbp]
   \centering
   \includegraphics[scale=.56,clip,trim=49mm 30mm 35mm 45mm,angle=0]{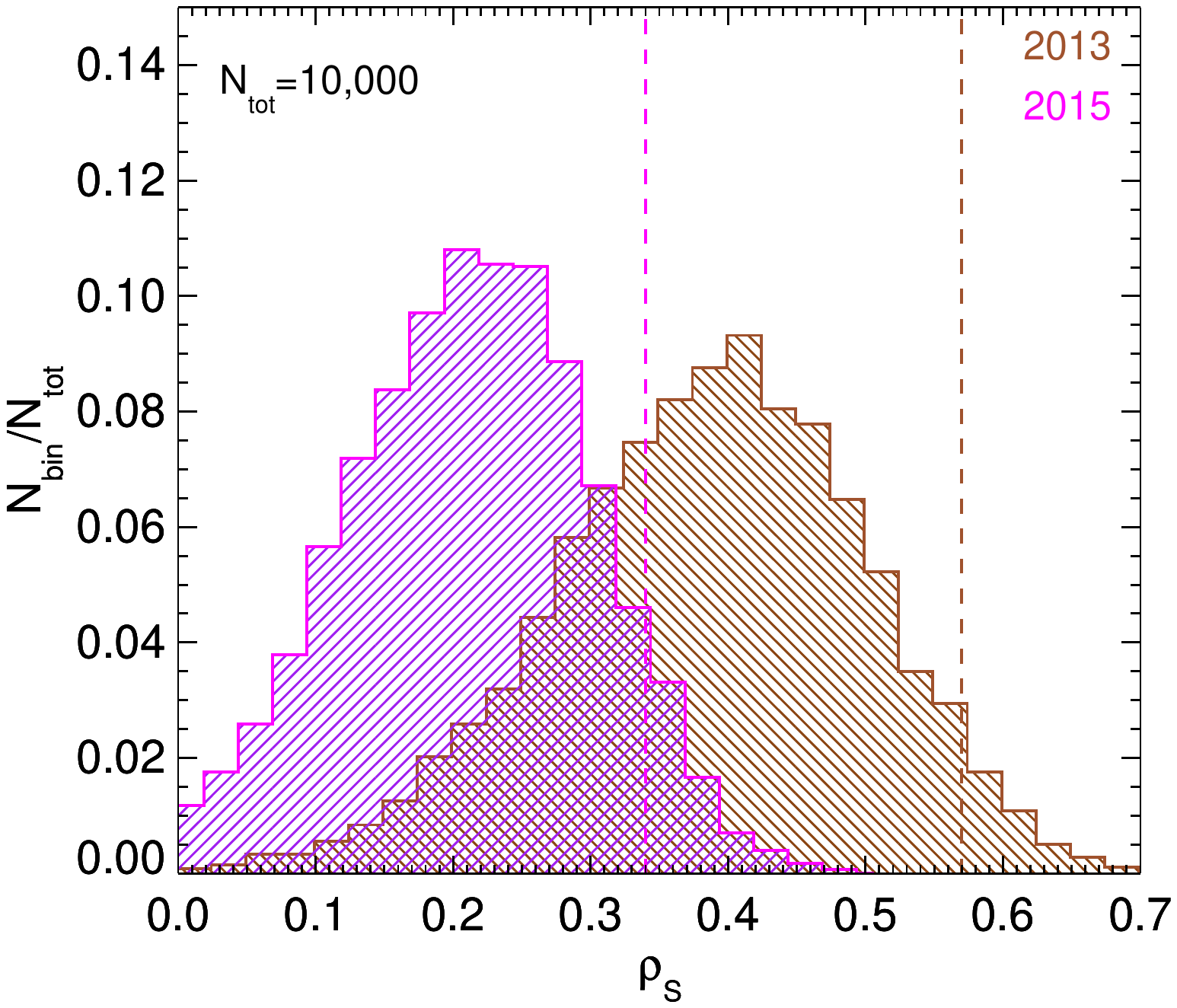} 
   \figcaption{Distributions of $\rho_s$ for the W$_{H\alpha}$ versus W$_{CaH}$ correlations shown in \autoref{fig:caii_corr}. 
   The distributions are generated by running 10,000 simulations and letting the values of W$_{CaH}$ and W$_{H\alpha}$
   vary according to their 1$\sigma$ uncertainties, which are assumed to be normal errors. The measured values
   of $\rho_S$ for both dates (vertical dashed lines) are much higher than the average values of the distributions,
   suggesting that the true correlations are weaker than the measured correlation.\label{fig:rhos_dist}}
\end{figure}

In order to test the significance of the correlations seen in both data sets, we have run a simple Monte Carlo
procedure that randomly draws values of \wca\ and \wha\ from a normal distribution defined by the 1$\sigma$
flux errors for each point. Representative values of the uncertainties are shown with the solid red lines in
\autoref{fig:caii_corr}. A simulation of 10,000 random draws is shown in \autoref{fig:rhos_dist} where the 2015
distribution of $\rho_S$ is shown in magenta and the 2013 distribution is shown in brown. The nominal $\rho_S$ values
calculated for the measurements in \autoref{fig:caiihatime_0704} and \autoref{fig:caiihatime_0804} are marked
with vertical dashed lines. It is clear that the nominal values lie on the high end of the distribution for both dates,
suggesting that the true correlation between \wca\ and \wha\ is weaker than the nominal values indicate. 

While we have focused on stellar activity being the main driver of the \ion{Ca}{2} core flux, it is also possible
that the changes in the core flux are in part due to absorption in the pre-transit material. This would
weaken any interpretation of the $W_{CaII}$-$W_{H\alpha}$ correlation as testing the relationship
between stellar activity in the two lines. On the other hand, any absorption by pre-transit material
in one line but not the other has a similar effect. Thus interpretation of the \ion{Ca}{2} residual
spectrum will also benefit from the activity baseline observations suggested for H$\alpha$ earlier in
this section. 

\subsection{The impact of stellar activity}
\label{sec:activityimpact}

The analysis presented in \autoref{sec:calcium2} suggests that the impact of stellar activity on the
Balmer line transmission signal is non-negligible and needs to be taken into account. The
correlation between \wca\ and \wha, although modest, weakens the interpretation of the pre-transit
signal presented in \citetalias{cauley15} as being caused by material compressed in a bow shock as
opposed to variations in the stellar activity level. Due to the level of scatter in the \wca\
values, and also the fact that \ion{Ca}{2} and H$\alpha$ do not track exactly the same physical
conditions or locations in the stellar chromosphere, we cannot empirically separate the contribution
of normal stellar activity from the observed W$_\lambda$ time series. 

\begin{figure}[htbp]
   \centering
   \includegraphics[scale=.53,clip,trim=45mm 30mm 35mm 45mm,angle=0]{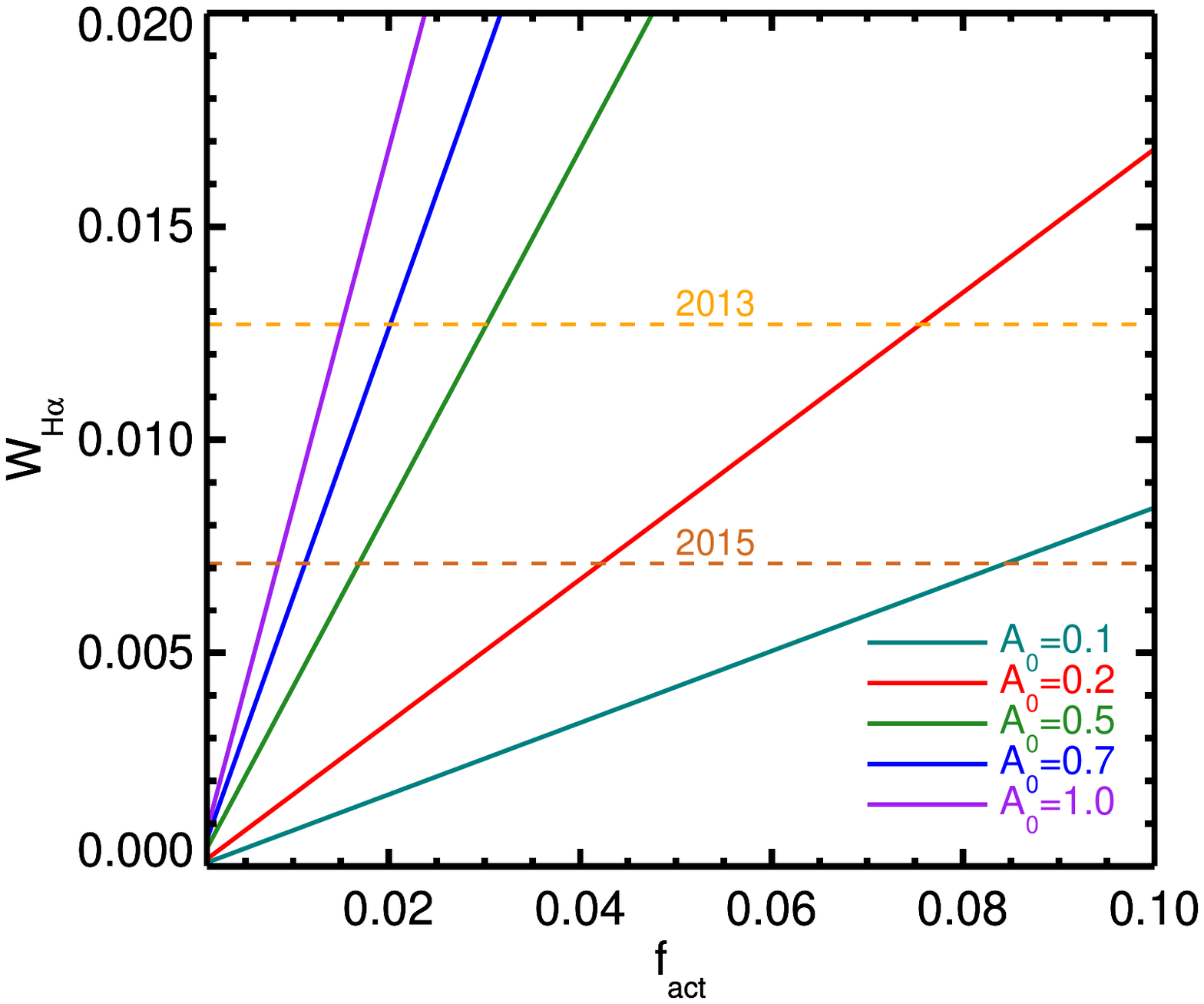} 
   \figcaption{W$_\lambda$ as a function of active region surface area fraction ($f_{act}$) for various normalized emission line strengths ($A_0$). 
   W$_\lambda$ is computed according to \autoref{eq:wlambda} for the transmission spectra defined in \autoref{eq:stactivity}.
   The horizontal dashed brown line is the mean value of W$_{H\alpha}$ for the pre-transit absorption values in 
   \autoref{fig:mabs_all}; the dashed orange line is the mean value of the first three pre-transit points from
   \citetalias{cauley15}. While the strength of the emission line is degenerate with f$_{act}$, it is clear that for
   a range of reasonable parameters the measured change in the pre-transit values of W$_\lambda$ for both the 2013 and 2015 transits
   can be accounted for by low levels of stellar activity.\label{fig:acttests}}
\end{figure}

In lieu of an empirical method for separating the stellar activity level from any true absorption, we can explore
the parameter space of active region surface coverage and emission line intensity required to reproduce
the observed changes in W$_\lambda$. We take an approach similar to that of \citet{berta} and examine
differences between weighted combinations of active region spectra, i.e., emission lines, and quiet star
spectra, in this case a flat continuum. The emission lines are Gaussians with FWHM $\sigma$ and height
$A_0$. We focus on H$\alpha$ and take $\sigma$=0.79 \AA, the mean FWHM of the pre-transit
absorption lines. The height $A_0$ is varied from 0.1 to 1.0. Low values of the emission line heights are
chosen to mimic the low levels of variation that we see here. Moderate solar and stellar flares can cause increases
in the emission line flux by factors of $\sim$2 relative to the local continuum \citep{johnskrull97,kowalski13} and
we do not see evidence (e.g., sharp increases in line strength followed by gradual decay) for large flaring
events in our spectra. In addition, large emission line strengths are not necessary to reproduce the observed
values of W$_{H\alpha}$, as shown below. 

Active region surface area coverage of the observable hemisphere is varied
from 0.1\% up to 10\% \citep[e.g.,][]{andretta95,meunier09}. Spot coverage for HD 189733 has been
estimated at $\sim$1\% \citep{winn07}. For the Sun, \citet{shapiro15} showed that the disk spot coverage fraction ($A_S$),
during peaks in the solar activity cycle, is $\sim$0.2\% while the disk active region coverage fraction ($A_F$) is
$\sim$3\%. They also showed that the ratio of $A_S$/$A_F$ increases as the level of solar activity increases.
This suggests that for HD 189733 the ratio $A_S$/$A_F$ is greater than that for the sun, although the
precise relationship is uncertain for very active stars. For the Sun during an activity maximum, $A_S$/$A_F$=0.07.
Thus we can assume with some confidence that the maximum active region coverage fraction of
HD 189733 b is less than 15\%. Furthermore, \citet{shapiro15} showed that brightness changes
in more active stars are dominated by spots rather than faculae (i.e., bright active regions), although
this does not rule out active region variability as the cause of small changes in spectral line flux. 

The transmission spectrum is calculated as

\begin{equation}\label{eq:stactivity}
S_T = \frac{S_* (1-f_{act}) + f_{act} S_{act}}{S_*} 
\end{equation}

\noindent where $f_{act}$ is the fraction of the observable hemisphere covered by active regions with the
corresponding spectrum $S_{act}$. The quiet star spectrum is $S_*$, which can be considered to be the
comparison spectrum. W$_{H\alpha}$ is then calculated according to \autoref{eq:wlambda}.  

\autoref{fig:acttests} shows the dependence of W$_{H\alpha}$ on the active region surface coverage and the
strength of the active region emission lines. The average observed value of W$_{H\alpha}$ for both dates is shown with a
 horizontal dashed line. The strength of the active region emission line, $A_0$, is degenerate with $f_{act}$. 
This results in many different combinations of $A_0$ and $f_{act}$ being able to reproduce the observed
changes in W$_{H\alpha}$. Regardless of the specific values of $A_0$ and $f_{act}$,
the measured levels of pre- and in-transit absorption in both the 2013 and 2015 data can be
approximated by relatively small changes in the stellar activity level between the comparison spectra and the spectrum
of interest. The smaller pre-transit values for the 2015 transit
can be reproduced by almost any combination of $f_{act}$ and $A_0$. The large values from the 2013 transit
can be reproduced by most of the parameter combinations but weak line strengths require large surface
coverage fractions. Although HD 189733 is an active star, it is unclear whether 8-10\% of the visible
hemisphere can be covered in bright active regions. High-cadence monitoring of HD 189733 while
HD 189733 b is not transiting will provide a better statistical understanding of activity variations
on $\sim$5 minutes timescales and help determine the true nature of the pre-transit signal.

For in-transit measurements, contrast between the comparison spectra, which are integrated across
the entire stellar disk, and the in-transit spectrum of interest, which is integrated across the
stellar disk not covered by the opaque planet, can result in line strength differences as the
spectrum is weighted towards active or spotted regions of the star \citep{berta}. The contrast
effect can be demonstrated by a planet with no atmosphere transiting a chord that is free of spots
or active-regions.  In this case, the in-transit spectrum will be weighted towards the spotted or
active stellar surface by the ratio ($R_p$/$R_*$)$^2$. For the HD 189733 system,
($R_p$/$R_*$)$^2$=0.024. If the out-of-transit spectrum is composed of 1.00\% spotted/active region
spectrum, the weighting of the in-transit spectrum, ignoring limb darkening, will change to 1.02\%
spotted/active and 98.98\% quiescent. Thus any difference between the spotted/active region spectrum
and the quiescent stellar spectrum will affect the transmission spectrum at the level of 0.02\%.
This is much smaller than the observed in-transit line depth. We note that this effect changes
weakly with increased spot/active region coverage and is $<$1\% even for coverage fractions of 30\%.
In addition, the in-transit signal begins before the optical transit and continues after the opaque
planetary disk leaves the stellar disk. This is further evidence that the absorption is caused by
gas in the planet's atmosphere and not the contrast effect since the contrast effect requires some
of the stellar surface to be blocked by the planet. 

\subsection{The choice of comparison spectra}
\label{sec:compspec} 

The analysis in \autoref{sec:activityimpact} shows that differing levels of stellar activity at different times
during the period of observation can affect the relative level of measured absorption in the line core. This
is directly related to the choice of comparison spectra. If instead of choosing spectra numbers 2-9 for the 2015 transit we select the
observations between $t-t_{mid}=-230$ and $-160$ minutes as the comparison spectra, all of the measured
W$_\lambda$ (\autoref{fig:mabs_all}) values would shift up by the relative average difference between the points in those time ranges. 
Thus instead of pre-transit \textit{absorption} we would observe an initial period of enhanced
stellar activity, followed by a relative period of quiescence, then another brief activity increase between
$t-t_{mid}=-150$ and $-116$ minutes, and finally relative quiescence immediately before the transit. If
instead we chose the post-transit spectra, as we chose to do in \citetalias{cauley15}, the in-transit
absorption would be very weak and all of the pre-transit W$_\lambda$ measurements would be positive,
suggesting sustained levels of higher stellar activity compared to in- and post-transit times. Thus clearly
the choice of comparison spectra can influence the interpretation of the W$_\lambda$ values. 

Our main justification for choosing the earliest observations in the 2015 data as the comparison spectra
is that they are located furthest from the transit and thus have a lower likelihood of being contaminated
by high-density pre-transit material capable of producing an absorption signature. It would not be unreasonable,
however, to select a different set of pre-transit spectra. We believe the low post-transit W$_\lambda$ values, and
their significant variability, rule these spectra out as good out-of-transit baseline values. The up-and-down
nature of the 2015 pre-transit W$_\lambda$ values do not offer much evidence for or against absorbing
material versus changes in the stellar activity level as the cause of the signal. However, if the large pre-transit
variations of $\sim$0.005-0.007 \AA\ at $-230$, $-155$, and $-110$ minutes were caused by changes in the stellar activity level, this
would indicate such variations are common and we might expect to see similar changes on similar timescales
during and after the transit. This is not the case: most of the in-transit variations are of smaller magnitude 
($\sim$0.003-0.005 \AA) and shorter timescales, which may be due to transits of active regions. 
Furthermore, the duration of the dip in W$_{H\alpha}$ starting at $-220$ minutes is $\sim$80 minutes which
is very close to the transit duration of a ``narrow" feature. It would be
surprising if the star's activity level happened to abruptly change on similar timescales as the transit
duration. 

Finally, although the individual $W_{H\beta}$ values are marginal, if the pre-transit
spectra between $-230$ and $-155$ minutes or between $-110$ and $-60$ minutes are chosen as
the comparison spectra, the in-transit H$\beta$ absorption becomes much weaker or disappears entirely. 
Our results demonstrate the need for a longer baseline of short cadence observations in order to establish the 
true out-of-transit stellar activity baseline. 

\section{Modeling the pre-transit absorption}
\label{sec:model}

In \citetalias{cauley15} we modeled the pre-transit absorption as arising in a thin bow shock orbiting $\sim$13 $R_p$
ahead of the planet. While this simple geometric model was able to account for the absorption time series and
the strength of the measured absorption, the favored model parameters resulted in two relatively unlikely
conclusions: 1. A low stellar wind speed of $\sim$40 km s$^{-1}$ at the planet's orbital radius; 2. A very strong
equatorial surface planetary magnetic field strength of $\sim$28 G. The low stellar wind speed is at odds with
most MHD simulations of the slow solar and stellar winds \citep[e.g.][]{cohen07,llama13,johnstone15} which reach speeds of
$\sim$200 km s$^{-1}$ at $\sim$10 $R_*$. We note, however, that observational constraints on the stellar winds of low mass
stars besides the Sun are weak. The large magnetic field strength from our model, derived assuming pressure balance
between the incoming stellar wind and planetary magnetosphere, is $\sim$4 times larger\footnote{In \citetalias{cauley15} the 
magnetic field strength was given as $\sim$2 times stronger than the \citet{reiners10} estimate. This was incorrect since the 
\citet{reiners10} given value is 14 G at the pole and not the equator. For a dipolar field, the equatorial value is half of
the polar value.} than that estimated by \citet{reiners10} for HD 189733 b based on magnetic field strength scaling relations 
for brown dwarfs and giant planets. While field strengths of this magnitude are estimated for more massive planets, it is difficult to produce
such a strong field for a Jupiter-mass object \citep{reiners09,reiners10}.

The absorption time series shown in \autoref{fig:mabs_all} does not match the specific prediction of the bow shock
model from \citetalias{cauley15}. Furthermore, we were unsuccessful at finding a suitable bow shock model
to describe the 2015 time series. As demonstrated in \autoref{sec:activityimpact}, the magnitude of the pre-transit
absorption can be influenced by low level variations in the stellar activity level. If the observed signal is assumed to
be solely associated with the planet, it is possible to explain the absorption as arising in transiting circumplanetary
material. This motivated us to search for a geometry that is capable of reproducing the absorption. 

\subsection{Clumpy accretion}
\label{subsec:clump}

One such geometry is an accretion stream escaping from the L1 point as the planet overflows
its Roche lobe \citep{lai,li10}. The relatively constant pre-transit W$_\lambda$ absorption values, the very early appearance
of the absorption, and the duration of the pre-transit signals are suggestive of a clumpy accretion flow. 
This type of accretion flow, i.e., non-uniform and time variable spiraling in from the planet's orbit, 
is seen in specific 2D MHD simulations of \citep[][; see also \citet{bisikalo13} for simulations of Roche lobe overflow
where the accretion stream is halted by the stellar wind pressure]{matsakos}. 

We have modeled the pre-transit absorption as being caused by two distinct clumps of material that are
spiraling in from the planet's orbital radius towards the central star. The choice of two clumps is 
motivated by the two distinct pre-transit absorption signatures seen in the W$_{H\alpha}$ values in \autoref{fig:mabs_all}.
The spiral trajectory in this case is approximated as a straight line since the clumps are located relatively close to the planet.
In fact, due to the combination of the Coriolis force and the gravitational attraction of the
planet, the trajectory of the material near the L1 point can be approximated by a straight line
\citep{lubow75,lai}. Any deviations from a straight line trajectory are secondary effects on the
transit light curve. Each clump is released from rest from the L1 point at 4.25 $R_p$ at some
time $t_{pre}$. The clump is then accelerated along a straight line trajectory that forms an angle
$\theta_c$ with the planet's orbit. The angle $\theta_c$ is determined by the system mass ratio and 
in this case is $\sim$60$^{\circ}$. The acceleration of the clump is $a_c$=0.0019 km
s$^{-2}$=2.3$\times$10$^{-8}$ $R_p$ s$^{-2}$, which is approximated from the
$q$=$M_p$/$M_*$=0.001 case ($q=$0.0013 for the HD 189733 system) presented in \citet{lai}. For simplicity, we
choose a cylindrical geometry for each clump defined by a radius $r_c$ and length $l_c$. The density
of each clump is uniform. The line profiles of the transiting clump are calculated identically to
the bow shock profiles in \citetalias{cauley15}. The optical depth at line center for each grid point is

\begin{equation}\label{eq:optdep}
\tau_0=\frac{\sqrt{\pi}e^2f_{lu}N_l\lambda_{lu}}{m_e c b} 
\end{equation}

\noindent where $f_{lu}$ is the oscillator strength of the transition, $\lambda_{lu}$ is the central wavelength, 
$m_e$ is the electron mass, $e$ is the electron charge, $c$ is the speed of light in vacuum, and $N_l$ is the column 
density of the lower energy level of the transition. The line profile is a Doppler-broadened delta function, $\tau_v=\tau_0 e^{-(v/b)^2}$, 
where $b=\sqrt{2}\sigma_v$ and $\sigma_v$ is the dispersion of a 1D Gaussian velocity distribution \citep{draine}.
The line broadening parameter for the clumps, used to approximate the profile shape according to 
\autoref{eq:optdep}, is $b$=5.7 km s$^{-1}$. 

\subsection{In-transit absorption}
\label{subsec:intran}

In order to maintain focus on the pre-transit signal, we will present a detailed discussion of the
in-transit data, including an analysis of velocities in the planetary atmosphere, in a subsequent
paper. A brief summary of the in-transit model, also shown over-plotted in \autoref{fig:mabs_all},
is given here. We model the 2015 in-transit absorption as arising in an extended uniform density
atmosphere. The choice of uniform density is motivated by the results of \citet{christie} who find
that the density of the dominant $n$=2 state (the 2s state) remains fairly constant over almost
three orders of magnitude in pressure (10$^{-6}$-10$^{-9}$ bar). This can be understood as the
result of outwardly increasing temperatures offsetting the decreasing neutral hydrogen abundance. We
assign a temperature to the base of the atmosphere and this sets the scale height, $H_{atm}$. The 3D
density grid is filled with material out to 10$H_{atm}$ and the column density is calculated for
each atmospheric point that occults the star. We find a good approximation of the average in-transit
Balmer line profile shape and absorption values with a density $\rho=4.0\times10^{-23}$ g cm$^{-3}$
and $T=6,000$ K, which translates to a scale height of $H_{exo}$=0.028 $R_p$.

\subsection{Model results}
\label{subsec:modresults}

When finding a model that roughly describes the pre-transit signal, we need to consider the strength
and shape of the $W_\lambda$ time series, the ratios of the Balmer line $W_\lambda$ values, and the
shape of the observed line profiles.  While these constraints taken together significantly narrow
the available parameter space, degeneracies exist between the model parameters. For example,
increasing the length of the clump by 50\% while decreasing the density by a factor of $\sim$3
results in similar average $W_\lambda$ values. The increase in clump length, however, lengthens the
egress and ingress time of the pre-transit features, which appear to be very brief ($\sim$10
minutes). In addition, the decreased density produces a narrower line profile as opacity broadening
becomes less prominent. This results in less acceptable line profile matches between the data and
the model. The final clump model values in \autoref{tab:modpars} reflect our attempt to balance
these constraints. 

%\capstartfalse           
\begin{deluxetable}{lccc}
%\rotate
%\linespacing{1}
\tablecaption{Model parameters  \label{tab:modpars}}
\tablehead{\colhead{Parameter}&\colhead{Symbol}&\colhead{Value$^{a,b}$}&\colhead{Units}\\
\colhead{(1)}&\colhead{(2)}&\colhead{(3)}&\colhead{(4)}}
\tabletypesize{\scriptsize}
\tablewidth{2000pt}
\startdata
Stellar radius & $R_*$ & 0.756 & $R_\Sun$\\
Stellar rotational  & $v$sin$i$ & 3.10 & km s$^{-1}$ \\
Impact parameters & $b$ & 0.680 & $R_*$ \\
Orbital period & $P_{orb}$ & 2.218573 & days \\
Orbital velocity & $v_{orb}$ & 151.96 & km s$^{-1}$  \\
Orbital radius & $a$ & 0.03099 & AU \\
Planetary radius & $R_p$ & 1.138 & $R_J$ \\
 & & &  \\
Time of clump initiation & $t_{pre}$ & 16.7,11.4 & hours before $t=0$  \\
Clump radius & $r_c$ & 0.5,0.5 & $R_p$  \\
Clump length & $l_c$ & 1.0,0.5 & $R_p$  \\
Clump density & $\rho_c$ & 1.5,4.0 & 10$^{-22}$ g cm$^{-3}$  \\
Line broadening & $b_c$ & 5.7,5.7 & km s$^{-1}$  \\
\enddata
\tablenotetext{a}{With the exception of $v$sin$i$, all stellar and planetary parameters taken from \citet{torres08}. The 
$v$sin$i$ value is taken from \citet{collier10}.}
\tablenotetext{b}{Clump parameters listed for clump 1 and then clump 2.}
\end{deluxetable}
%\capstarttrue

\begin{figure}[htbp]
   \centering
   \includegraphics[scale=.51,clip,trim=10mm 15mm 75mm 10mm,angle=0]{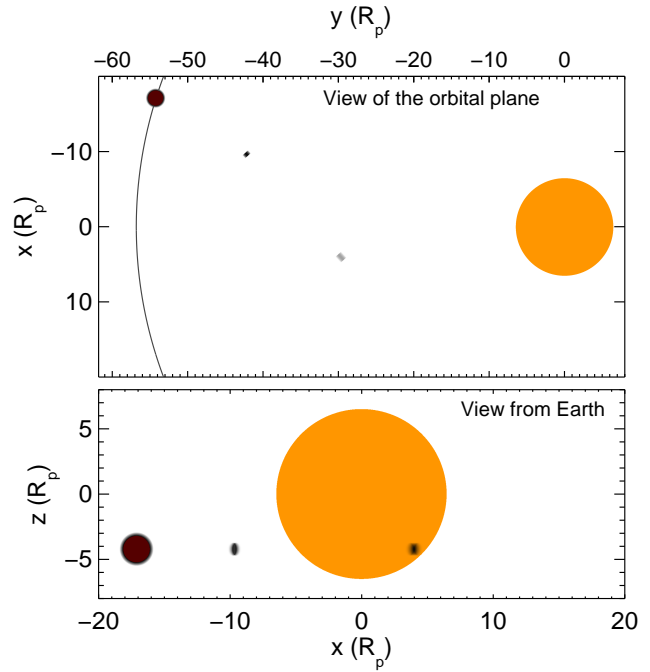} 
   \figcaption{Views of the first transiting clump in the plane of the orbit (top panel) and perpendicular to the orbital plane (bottom
   panel) for the snapshot at $t-t_{mid}=-158$ minutes. The densities are scaled to the planetary atmosphere. The clumps 
   represent stochastic mass loss events from the planet that are dense and extended enough to cause significant absorption
   in the cores of the Balmer lines.\label{fig:tranview}}
\end{figure}

The model W$_\lambda$ values are shown as solid lines in \autoref{fig:mabs_all}. The final clump
parameters and the systems parameters used in the model are listed in \autoref{tab:modpars}. A view
of the clump transit is shown in \autoref{fig:tranview} for $t-t_{mid}=-158$ minutes\footnote{An animation
of the clump transit is available at pwcastro.com/research/.}. The clump
model is successful at describing both the timing of the pre-transit events and their absorption
levels. We note that if the clumps are not allowed to accelerate, the transit duration is too long
and the transit timing is significantly off. The densities of the clumps also roughly reproduce the
Balmer line absorption ratios and line shapes, although most of the individual H$\beta$ and
H$\gamma$ points are not measured at a significant level. 

One piece of evidence that could support the accretion clump interpretation is the large $S_{HK}$
value measured for the 2015 transit. If the star is in an active state, the planet will be subjected
to higher levels of high energy stellar radiation, increasing the mass loss rate \citep[e.g.,
][]{murray09,trammell,owen,valsecchi15,owen16}. If enough material is evaporated from the atmosphere, the planet's
magnetosphere can limit the outflow rate if the magnetic pressure dominates the gas pressure,
creating a ``dead zone'' that extends many planetary radii \citep{trammell,trammell14,khodachenko}. Periodic
mass loss, i.e., the accretion clumps, is not implausible from a filled magnetosphere that is being
subjected to variable heating rates \citep[e.g., ][]{pillitteri15}. Furthermore, mass loss from HD 189733 b is observed to be
variable \citep{desetangs12}. Although we do not explore the in-transit signal in detail
here, the in-transit absorption appears to begin $\sim$15 minutes before $t_I$ and last at least 30
minutes after $t_{IV}$. The planet moves across the star at $\sim$0.11 $R_p$ minute$^{-1}$ which
means the feature causing the absorption immediately pre- and post-transit extends $\sim$1.5 $R_p$
ahead of the planet and $\sim$3.0 $R_p$ behind the planet. This extended cloud of material is
suggestive of a partially filled Roche lobe. 

\subsection{Limitations to the clumpy accretion interpretation}
\label{subsec:modlims}

While the accretion clump model can reproduce the observed absorption levels, and in addition to the
parameter degeneracies discussed above, there are serious limitations to this interpretation. The
first is the lack of large red-shifted velocities in the pre-transit H$\alpha$ transmission spectra
(\autoref{fig:tspecs_halpha} and \autoref{fig:halinevels}). By the time the clumps transit the star,
they are moving with line-of-sight velocities of $\sim$100 km s$^{-1}$ and $\sim$70 km s$^{-1}$,
much larger than the 5-10 km s$^{-1}$ line centroids seen for the individual spectra.  This is the
result of the large angle (60$^{\circ}$) that the clump trajectory makes with the planet's orbit.
This angle would need to be $\sim$5-10$^{\circ}$ in order to produce red-shifted velocities of 5-10
km s$^{-1}$. A possible resolution to this problem is if the trajectory of the accreting material is
altered by interactions with the stellar wind \citep[e.g., see ][]{matsakos}, although modeling
these interactions is beyond the scope of this paper. It is also not clear if simple hydrodynamic
Roche-lobe overflow, as we've assumed here, is directly applicable in the hot Jupiter regime,
especially when the confining effect of the planet's magnetosphere is taken into account
\citep{trammell,owen}. Thus a more realistic treatment of the mass loss could allow the trajectory
of the clumps to be less steep. 

Finally, there is currently little theoretical evidence to support the existence of large $n=2$
neutral hydrogen populations at such short distances from the star. During transit, the clumps are
within $\sim$5-6 $R_*$ of the star, about 30-40\% closer than the planet's orbit, and receive a
factor of $\sim$2$\times$ more stellar radiation. Again, the clump trajectory could be less extreme
through interactions with the stellar wind. Almost all of the escaping planetary material is
photo-ionized by $\sim$5 $R_p$ from the planet \citep[e.g.,][]{murray09,tripathi15,salz16} and the stellar
wind is entirely composed of ions. Thus some highly non-equilibrium process is needed to maintain a
population of excited neutral hydrogen for at least the duration of the transit. It is not clear if
such a process (e.g., time-dependent recombination or charge-exchange of stellar wind protons with
the small population of planetary neutrals) is capable of producing the necessary neutral hydrogen
density. Modeling of the interaction between the planetary accretion flow and the stellar wind and
how this interaction affects the excited neutral hydrogen density would be useful to clarify this
outstanding question. 

\section{Summary and conclusions}
\label{sec:summary}

We have presented new transit observations of HD 189733 b which cover a significant amount of
pre-transit phase and a small amount of post-transit phase. Our results can be summarized as the
following:

\begin{itemize}

\item We detect strong in-transit absorption from the planetary atmosphere in
H$\alpha$, H$\beta$, and H$\gamma$. We also detect weak \ion{Na}{1} 5896 \AA\ absorption and measure
a marginal amount of \ion{Mg}{1} 5184 \AA\ absorption. An upcoming paper will present an 
interpretation of the in-transit H$\alpha$ line centroid velocities. 

\item We confirm the presence of a pre-transit Balmer line signature at
a significant level for the 2015 data. However, the pre-transit signal observed in the 2015 transit
differs significantly from that observed in \citetalias{cauley15}. In particular, the 2015 pre-transit signal
does not match the prediction of the specific bow shock model presented in \citetalias{cauley15}. 

\item We have reanalyzed the \ion{Ca}{2} H line from the 2013 transit and found a moderate
correlation between the $W_{H\alpha}$ values and residual core flux of the \ion{Ca}{2} lines.  
A weaker correlation was found between
$W_{H\alpha}$ and $W_{CaII}$ for the 2015 data. We suggest that the residual core flux be used in
future studies as an activity proxy for short timescales rather than the $S_{HK}$ index. 

\item Motivated by the $W_{H\alpha}$-$W_{CaII}$
correlations, we explored the potential of changing activity levels to produce the pre-transit
signature. We find that even small levels of intrinsic stellar variation in the line cores can
reproduce the observed transmission spectra for both pre-transit measurements. It is not clear,
however, if changes in the stellar activity level on these timescales are common. 

\item We have modeled the pre-transit absorption as arising in a clumpy accretion flow. The clumpy
accretion model is able to reproduce both the timing and magnitude of the pre-transit events.  
However, there are significant drawbacks to the clumpy accretion model, the
foremost of which is the lack of large red-shifted absorption velocities that should be produced by
ballistically in-falling material. 

\end{itemize}

\added{Although we have modeled the pre-transit signature as arising in absorbing circumplanetary
material, we caution against a firm interpretation of the data. The magnitude of the pre-transit detections from both the 2013 and 
2015 data can be explained by changing levels of stellar activity. Due to the lack of detailed information
concerning short-term, low-level variations in the H$\alpha$ core flux, no conclusive argument can be 
made concerning one interpretation over the other. Both physical scenarios should be investigated for other hot planet
systems to give the current study context. We are currently pursuing a high-cadence H$\alpha$ monitoring
campaign of HD 189733, at all phases of the planet's orbit, with the intention of determining the frequency of 
these low-level changes in the core of the Balmer lines.}

Pre-transit signatures around hot planets may provide unique information concerning the dynamics of
escaping planetary material and, potentially, the structure and strength of the planetary
magnetosphere. Hot planets around less active 
stars than HD 189733 would make prime targets for future investigations into pre-transit signatures since 
the observed signal can be attributed to the planet with a higher probability. 

\bigskip

{\bf Acknowledgments:} We are grateful to the referee Joe Llama for comments that helped improve this manuscript.
The data presented herein were obtained at the W.M. Keck Observatory from
telescope time allocated to the National Aeronautics and Space Administration through the agency's
scientific partnership with the California Institute of Technology and the University of California.
This work was supported by a NASA Keck PI Data Award, administered by the NASA Exoplanet Science
Institute. The Observatory was made possible by the generous financial support of the W.M. Keck
Foundation. The authors wish to recognize and acknowledge the very significant cultural role and
reverence that the summit of Mauna Kea has always had within the indigenous Hawaiian community.  We
are most fortunate to have the opportunity to conduct observations from this mountain. This work was
completed with support from the National Science Foundation through Astronomy and Astrophysics
Research Grant AST-1313268 (PI: S.R.). A. G. J. is supported by NASA Exoplanet Research Program 
grant 14-XRP14\_2-0090 to the University of Nebraska-Kearney. 
P.W.C. is grateful for useful exchanges with T. Matsakos
concerning the morphology and dynamics of hot Jupiter accretion flows. P.W.C. and S. R. acknowledge
helpful conversations with D. Christie and C. Huang regarding the interpretation of neutral hydrogen
absorption in the extended planetary atmosphere. 

\clearpage

\appendix

\section{Individual transmission spectra}
\label{sec:tspecs_app}

\begin{figure*}[htbp]
   \centering
   \includegraphics[scale=.90,clip,trim=15mm 20mm 0mm 5mm,angle=0]{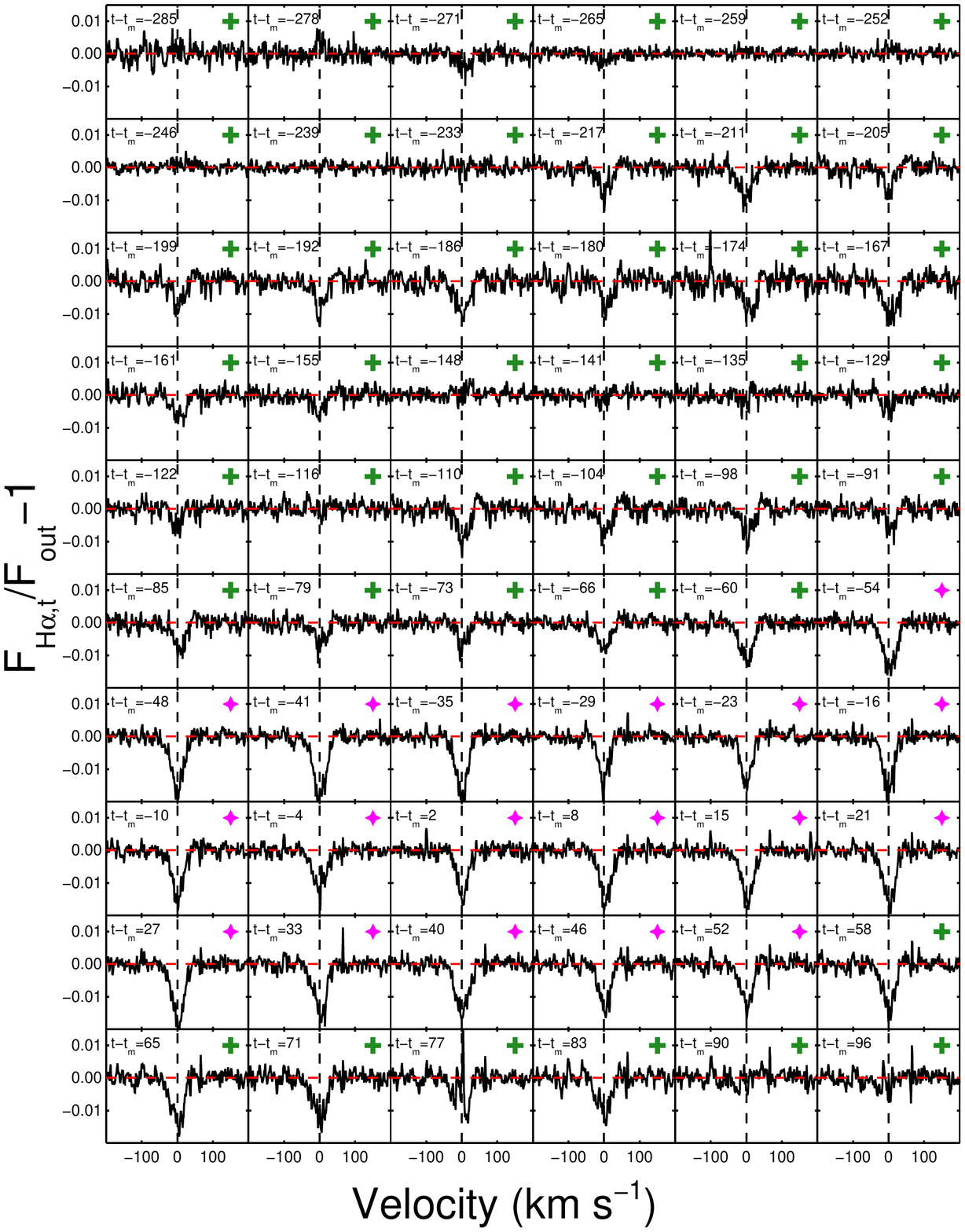} 
   \figcaption{Individual transmission spectra for H$\alpha$. Out-of-transit spectra are marked with a green cross
   and in-transit spectra are marked with a magenta star. The time from mid transit in minutes is labeled
   $t-t_{m}$.\label{fig:tspecs_halpha}}
\end{figure*}

\begin{figure*}[htbp]
   \centering
   \includegraphics[scale=.90,clip,trim=15mm 20mm 0mm 5mm,angle=0]{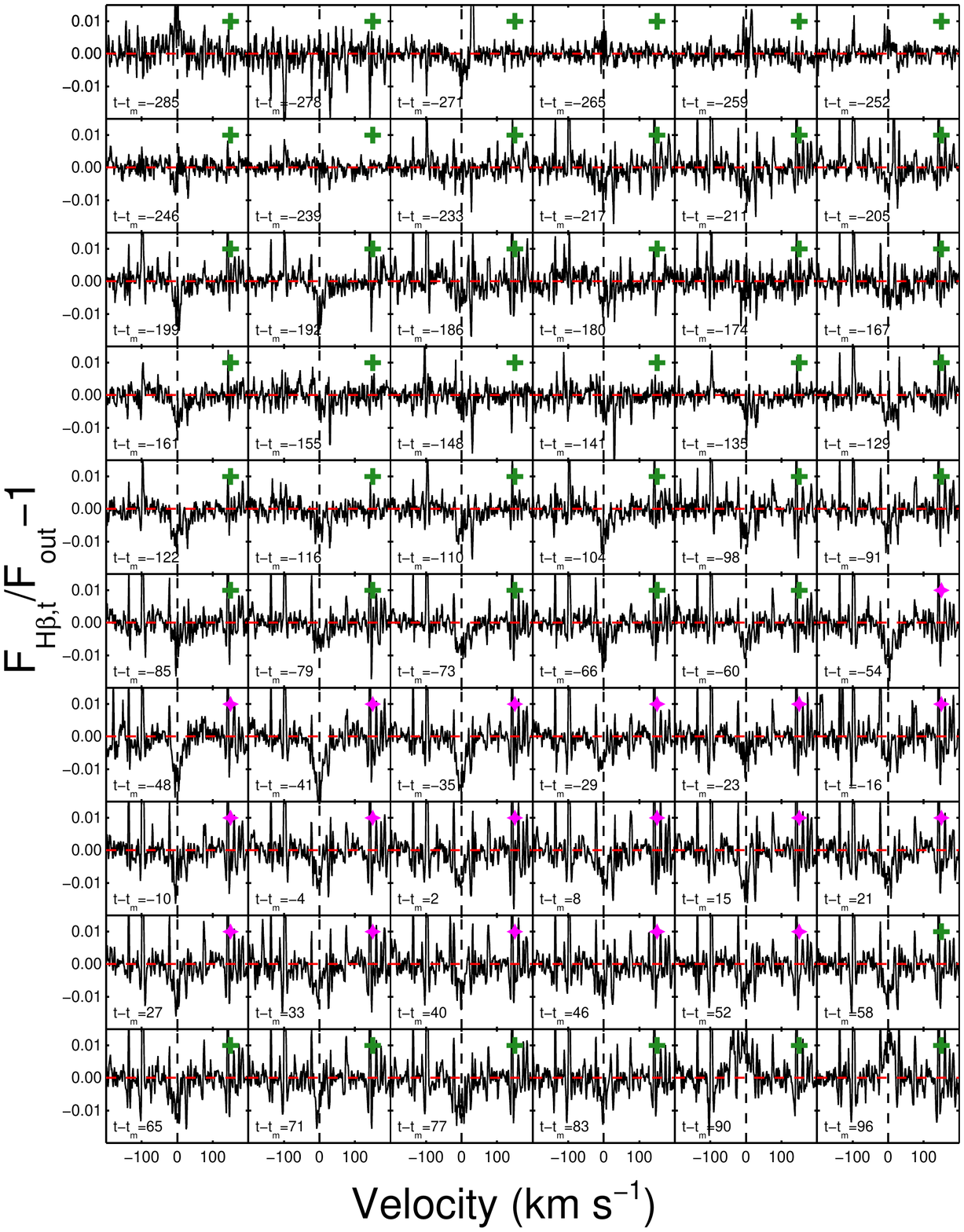} 
   \figcaption{Individual transmission spectra for H$\beta$. All markings are the same as \autoref{fig:tspecs_halpha}.\label{fig:tspecs_hbeta}}
\end{figure*}

\begin{figure*}[htbp]
   \centering
   \includegraphics[scale=.90,clip,trim=15mm 20mm 0mm 5mm,angle=0]{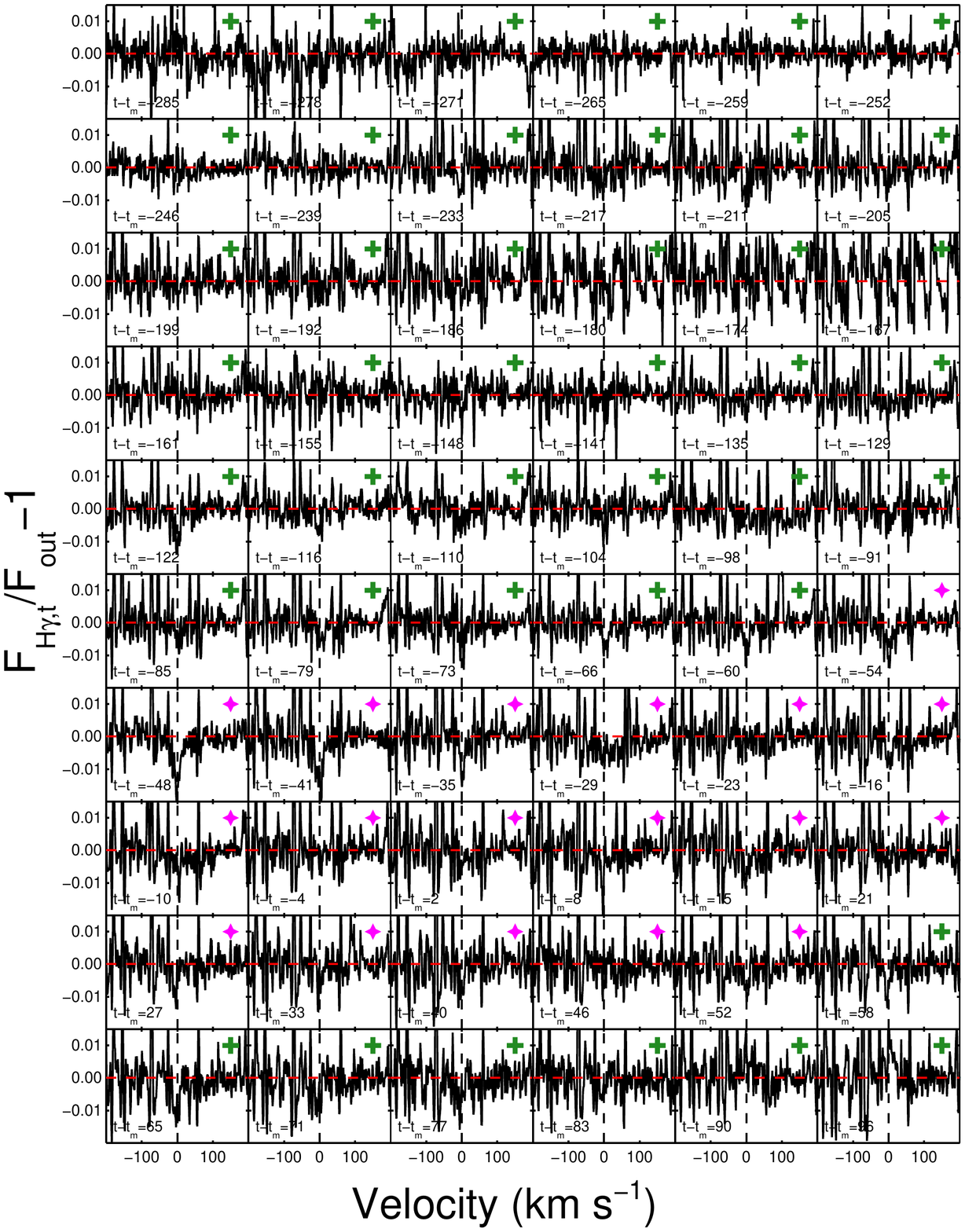} 
   \figcaption{Individual transmission spectra for H$\gamma$. All markings are the same as \autoref{fig:tspecs_halpha}
   and \autoref{fig:tspecs_hbeta}. The jagged sawtooth pattern in the spectra from $t-t_{m}=-$180 minutes to $-167$ minutes
   is likely caused by poor guiding and small amounts of order overlap on the blue chip.\label{fig:tspecs_hgamma}}
\end{figure*}

\begin{figure*}[htbp]
   \centering
   \includegraphics[scale=.90,clip,trim=15mm 20mm 0mm 5mm,angle=0]{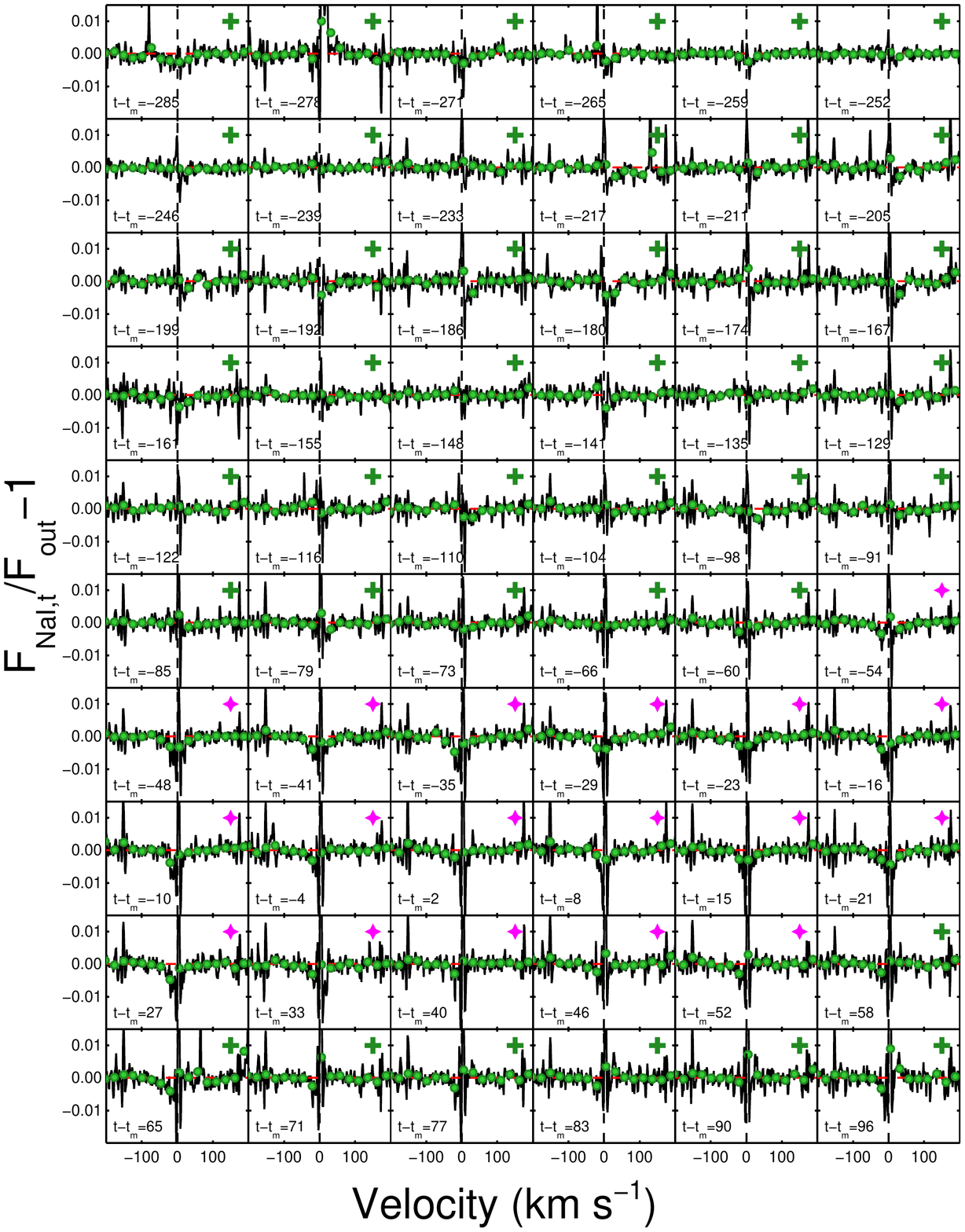} 
   \figcaption{Individual transmission spectra for \ion{Na}{1} 5896 \AA. The green dots show the spectra
   binned by 20 pixels for clarity. Note the weak red-shifted absorption feature in some of the pre-transit
   spectra.\label{fig:tspecs_nai}}
\end{figure*}

\begin{figure*}[htbp]
   \centering
   \includegraphics[scale=.90,clip,trim=15mm 20mm 0mm 5mm,angle=0]{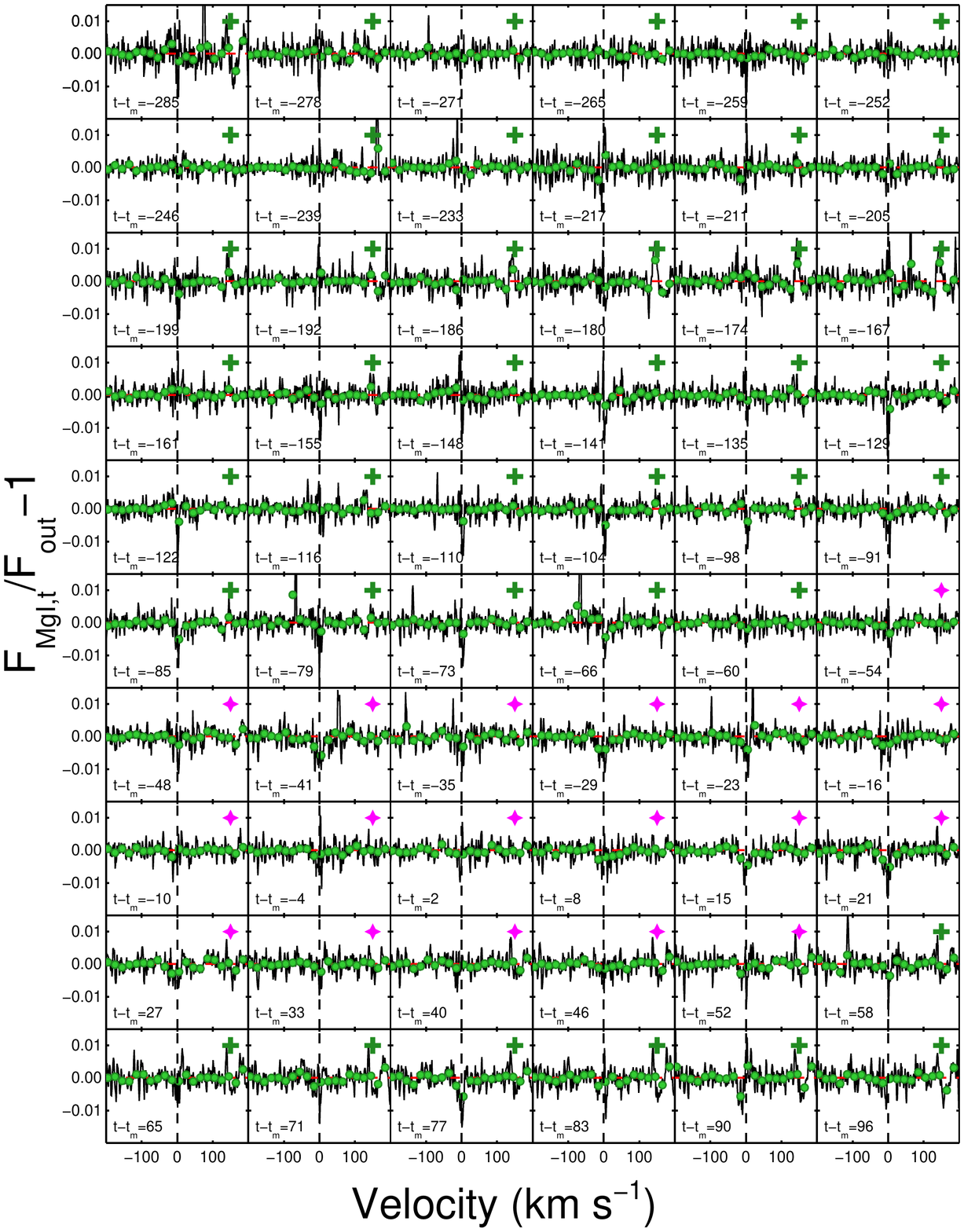} 
   \figcaption{Individual transmission spectra for \ion{Mg}{1} 5184 \AA. The green dots show the spectra
   binned by 20 pixels for clarity. The in-transit absorption, if present, is very weak and
   not detected at a significant level.\label{fig:tspecs_mgi}}
\end{figure*}

Transmission spectra for individual exposures are shown in \autoref{fig:tspecs_halpha}, \autoref{fig:tspecs_hbeta},
\autoref{fig:tspecs_hgamma}, \autoref{fig:tspecs_nai}, and \autoref{fig:tspecs_mgi}. The out-of-transit spectra are 
marked with green crosses; in-transit spectra are marked with a magenta star. The solid green dots in \autoref{fig:tspecs_nai} 
and \autoref{fig:tspecs_mgi} show the spectra binned by 20 pixels. The individual transmission spectra are the weighted
average of the ratio between the spectrum of interest and all combinations ($N$=255 for the eight comparison spectra) of the 
comparison spectra used to construct the master comparison spectrum. Transmission spectra for the individual comparison 
spectra are constructed by comparing the exposure of interest to a master comparison spectrum that includes all of the other comparison
spectra besides the spectrum of interest. This weighted average is performed to prevent individual comparison
spectra from having a large influence on the individual transmission spectra. 

Overall, the individual spectra are well defined for H$\alpha$ and for the strong H$\beta$ absorption. The individual
H$\gamma$ spectra, however, are very noisy and the absorption is not obvious except for the times $t-t_{mid}=-$54, $-48$,
and $-41$ minutes. We note that the telescope experienced minor guiding errors at $\sim$$-180$ minutes (see \autoref{fig:halphavcorrs}). On the
blue chip, where H$\gamma$ is located, the orders are closely spaced and any significant deviation of the stellar
profile can result in small amounts of order overlap. For these exposures, the stellar profile in the spatial
direction on the blue CCD is 10-15\% wider than it is for the other exposures, over which it is constant to within
$\sim$1-2\%. We believe this is what accounts for the sawtooth pattern seen
in the the H$\gamma$ spectra from $t-t_{m}=-$180 minutes to $-167$ minutes. Thus the absorption values
calculated from these spectra are highly suspect. The individual \ion{Na}{1} line profile morphologies are visible in a
few of the pre-transit and in-transit spectra in \autoref{fig:tspecs_nai}. However, large residuals in the line core
obscure the true shape of the line in most cases. The absorption, if present, is too weak to be seen in all but a few
of the \ion{Mg}{1} profiles in \autoref{fig:tspecs_mgi}.

\section{\ion{Ca}{2} residual core profiles}
\label{sec:caii_app}

\added{This appendix contains the \ion{Ca}{2} residual core profiles for both transit dates. Pre-transit profiles are shown in pink, in-transit profiles in purple,
and post transit profile in blue.}

\begin{figure*}[htbp]
   \centering
   \includegraphics[scale=.90,clip,trim=15mm 25mm 5mm 5mm,angle=0]{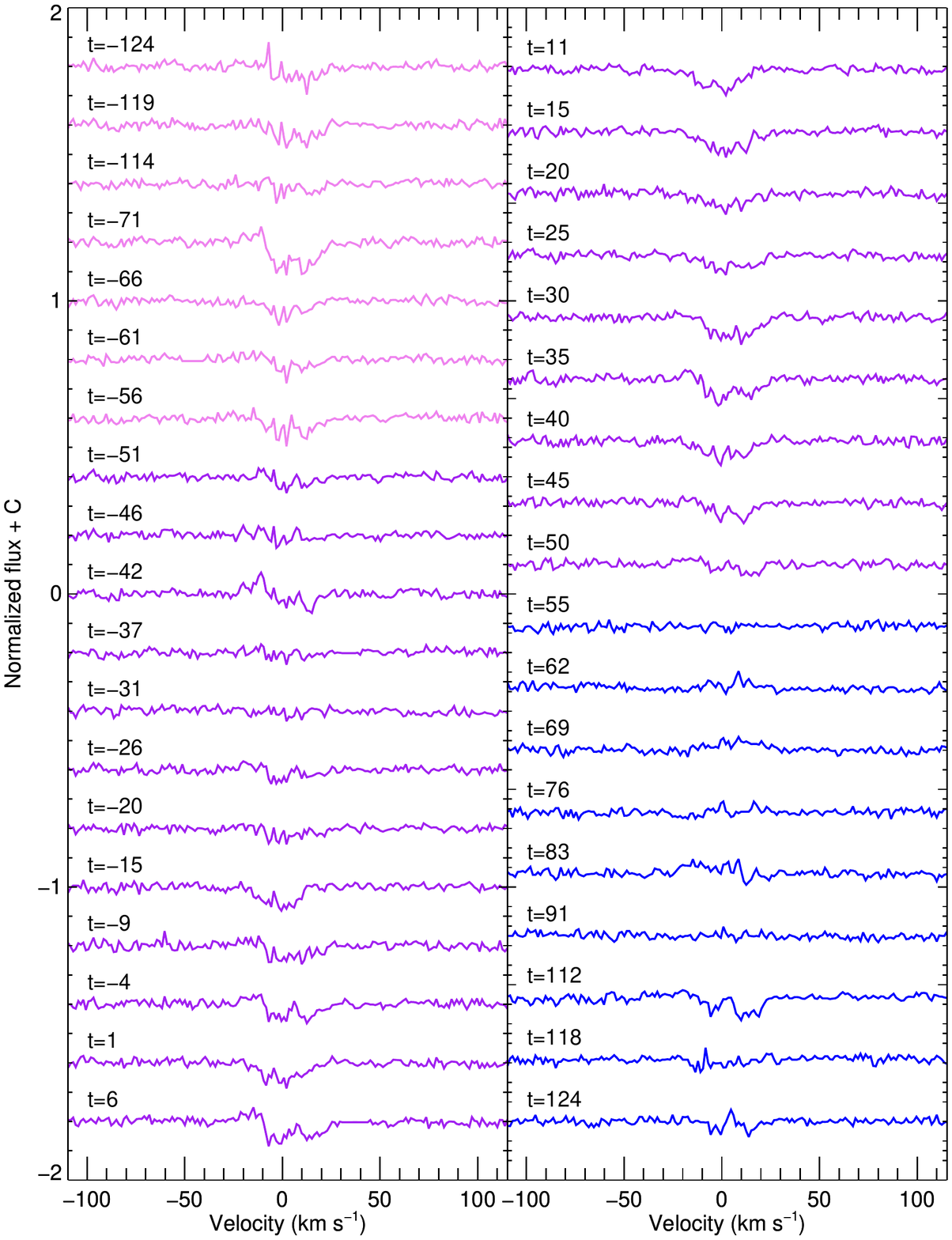} 
   \figcaption{Residual \ion{Ca}{2} H core profiles for the 2013 transit. The time given for each observation is in units of 
   minutes from mid-transit.\label{fig:caiires_0704}}
\end{figure*}

\begin{figure*}[htbp]
   \centering
   \includegraphics[scale=.90,clip,trim=15mm 25mm 5mm 5mm,angle=0]{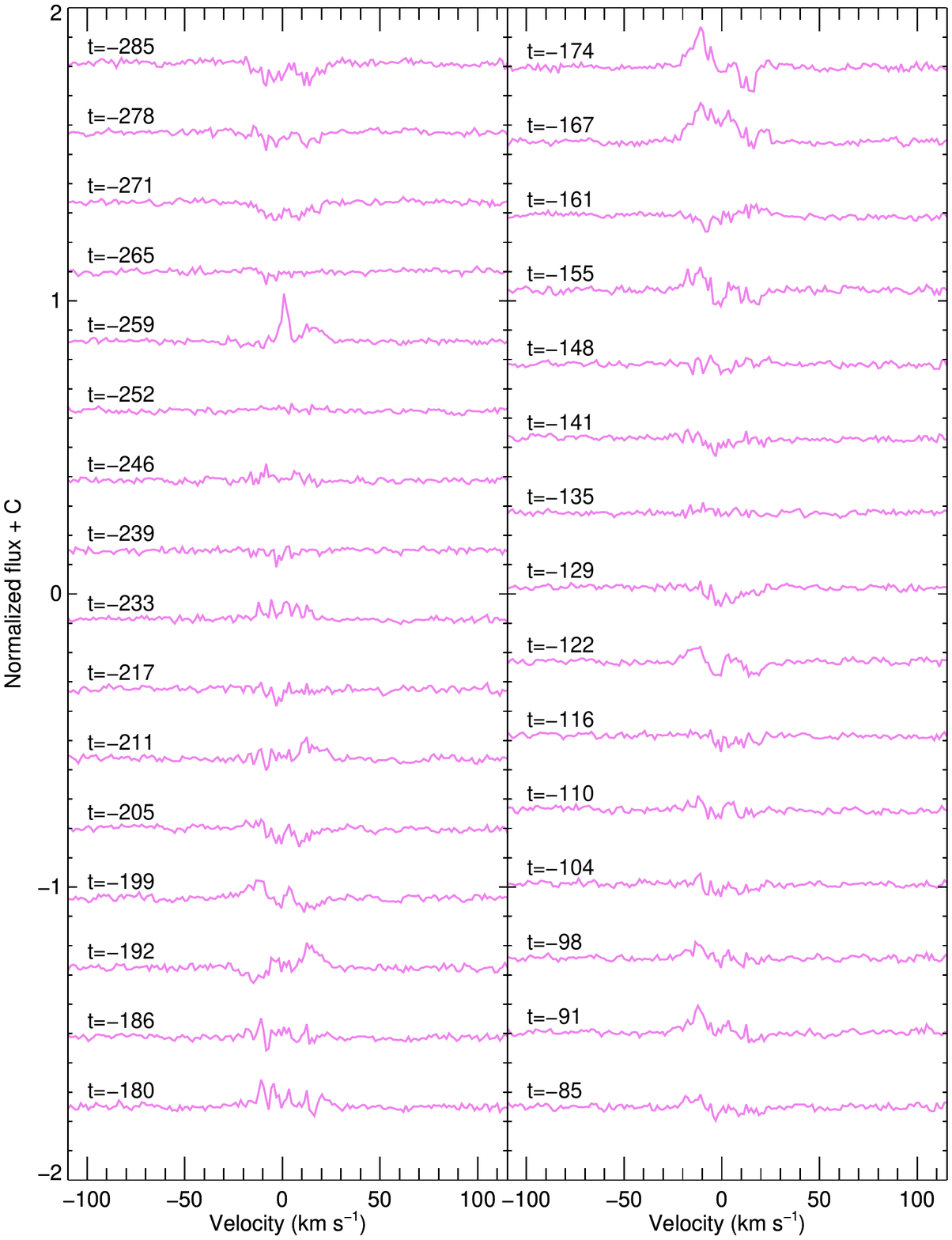} 
   \figcaption{Residual \ion{Ca}{2} H core profiles for the first part of the 2015 transit. Colors and labels are the same as \autoref{fig:caiires_0704}.
   \label{fig:caiires_0804_1}}
\end{figure*}

\begin{figure*}[htbp]
   \centering
   \includegraphics[scale=.90,clip,trim=15mm 25mm 5mm 5mm,angle=0]{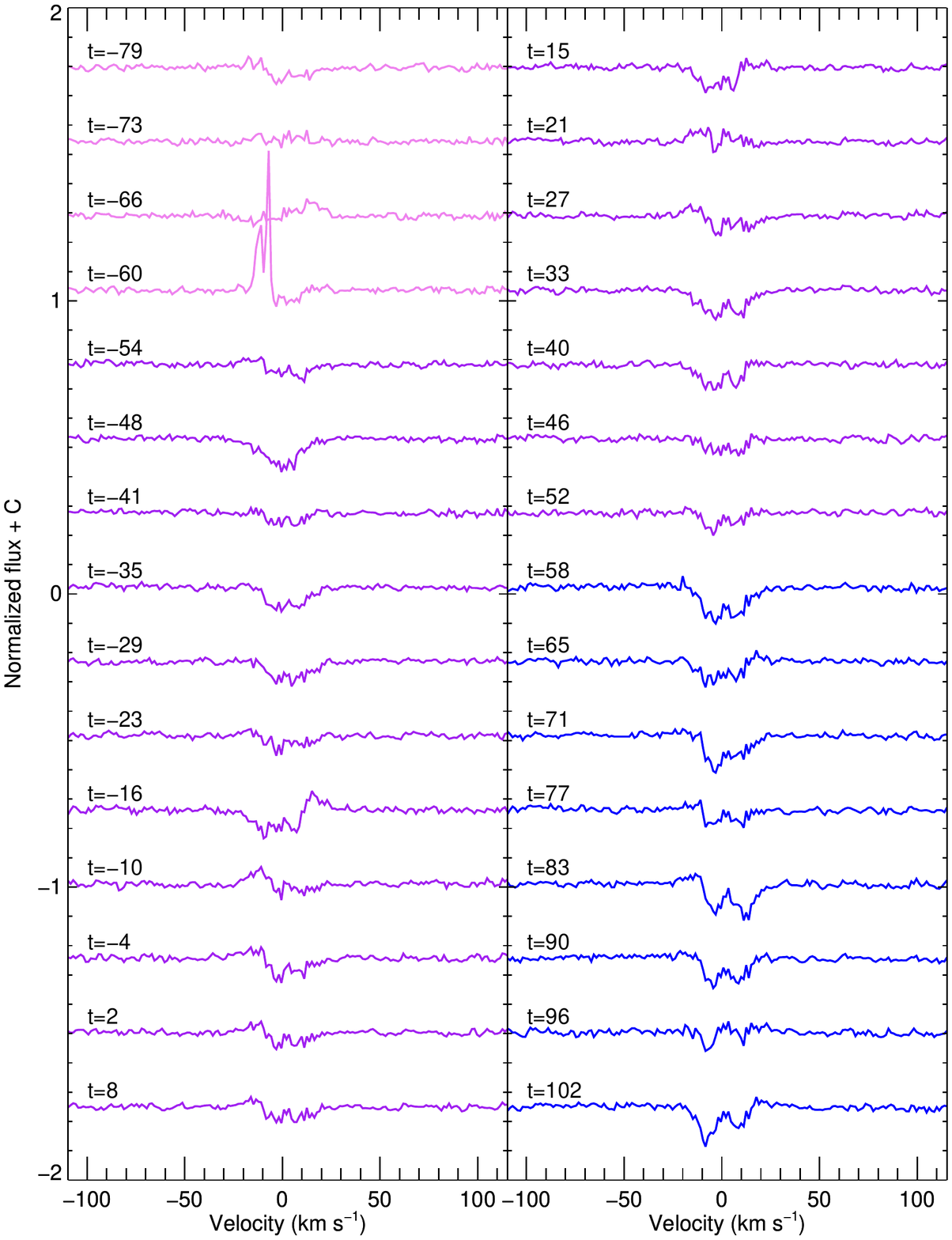} 
   \figcaption{Residual \ion{Ca}{2} H core profiles for the second part of the 2015 transit. Colors and labels are the same as \autoref{fig:caiires_0704}
   and \autoref{fig:caiires_0804_1}.\label{fig:caiires_0804_2}}
\end{figure*}

\end{document}